
%
%
\input harvmac.tex
\input epsf.tex
\newcount\figno
\figno=0
\def\fig#1#2#3#4{
\par\begingroup\parindent=0pt\leftskip=1cm\rightskip=1cm\parindent=0pt
\baselineskip=11pt
\global\advance\figno by 1
\midinsert
\epsfxsize=#3
\centerline{\epsfbox{#2}}
{\bf Fig.\ #4: } #1\par
\endinsert\endgroup\par
}
\def\figlabel#1{\xdef#1{\the\figno}}
\def\encadremath#1{\vbox{\hrule\hbox{\vrule\kern8pt\vbox{\kern8pt
\hbox{$\displaystyle #1$}\kern8pt}
\kern8pt\vrule}\hrule}}
%

\noblackbox
\lref\AFL{ N. Andrei, K. Furuya and J. Lowenstein, Rev.
Mod. Phys. 55 (1983) 331.}
\lref\pkondo{P. Fendley, Phys. Rev. Lett. 71 (1993) 2485,
cond-mat/9304031.}
\lref\FLSbig{P. Fendley, A. Ludwig, H. Saleur,
cond-mat/9503172, to appear in Phys. Rev. B.}
\lref\FSkon{P. Fendley and H. Saleur,
``Exact perturbative solution of the Kondo problem'',
cond-mat/9506104.}
\lref\FLSnoise{P. Fendley, A. Ludwig and H. Saleur,
Phys. Rev. Lett. 75 (1995) 2196, cond-mat/9505031.}
\lref\FSW{P. Fendley, H. Saleur and N.P. Warner,
Nucl. Phys. B 430 (1994) 577, hep-th/9406125.}
\lref\YA{ P.W. Anderson and G. Yuval, Phys. Rev. Lett.
23 (1969) 89;
G. Yuval and P.W. Anderson, Phys. Rev. B1 (1970) 1522.
}\lref\Schotte{ K.-D. Schotte, Z. Physik 230 (1970) 99.}
\lref\TW{ A.M. Tsvelick and P.B. Wiegmann, Adv. Phys. 32 (1983)
453.
}\lref\Kondo{ J. Kondo, Prog. Th. Phys. 32 (1964) 37.
}\lref\FLSjack{P. Fendley, F. Lesage and H. Saleur,
J. Stat. Phys. 79 (1995) 799, hep-th/9409176.
}\lref\BLZ{ V. Bazhanov, S. Lukyanov and A.B. Zamolodchikov,
``Integrable Structure of Conformal Field Theory, Quantum KdV
Theory, and Thermodynamic Bethe Ansatz,'' hep-th/9412229.}
\lref\BLZnew{ V. Bazhanov, S. Lukyanov and A.B. Zamolodchikov,
to appear.}
\lref\Leggett{ A.J. Leggett, S. Chakravarty, A.T. Dorsey,
M.P.A. Fisher, A. Garg and W. Zwerger, Rev. Mod. Phys. 59
(1987) 1. For subsequent developments, see S. Chakravarty and
J. Rudnick, Phys. Rev. Lett. 75 (1995) 701.
}\lref\KF{C.L. Kane and M.P.A. Fisher, Phys. Rev. B46 (1992)
15233.
}\lref\FLS{P. Fendley, A.W.W. Ludwig and H. Saleur, Phys.
Rev. Lett. 74 (1995) 3005, cond-mat/9408068.}
\lref\periodtba{Al.B. Zamolodchikov,
Phys. Lett. B253 (1991) 391; E. Frenkel and A. Szenes,
``Thermodynamic Bethe Ansatz and Dilogarithm Identities I'',
hep-th/9506215.}
\lref\tsvel{A.M. Tsvelik, ``Existence of Low-Temperature Critical
Regime in 1D Luttinger Liquid with a Weak Link'',
cond-mat/9502022.}
\lref\lem{K. Leung, R. Egger and C. Mak,
``Dynamical simulation of transport in one-dimensional quantum
wires'',
cond-mat/9509078.}
\lref\exper{F.P. Milliken, C.P. Umbach and R.A. Webb,
``Indications of a Luttinger Liquid in the Fractional Quantum Hall
Regime'', to be published.}
\lref\Moon{ K. Moon, H. Yi,  C.L. Kane,
S.M. Girvin and M.P.A. Fisher, Phys. Rev. Lett. 71 (1993) 4381,
cond-mat/9304010.}
\lref\GZ{S. Ghoshal and A.B. Zamolodchikov, Int. J. Mod.
Phys. A9 (1994) 3841, hep-th/9306002.}
\lref\Wen{X.G. Wen, Phys. Rev. B41 (1990) 12838;
Phys. Rev. B43 (1991) 11025.}
\lref\Tracy{C. Tracy and H. Widom,
``Proofs of Two Conjectures Related to the Thermodynamic Bethe
Ansatz'',
solv-int@xyz.lanl.gov/9509003.}

\Title{USC-95-20}
{\vbox{
\centerline{A unified framework for the Kondo problem}
 \vskip 4pt
\centerline{and for an impurity
in a Luttinger liquid}}}

\centerline{P. Fendley, F. Lesage and H. Saleur\footnote{$^\dagger$}
{Packard Fellow}}
\bigskip\centerline{Department of Physics}
\centerline{University of Southern California}
\centerline{Los Angeles, CA 90089-0484}

\vskip .3in

We develop a unified theoretical
framework for the anisotropic Kondo model and the boundary
sine-Gordon model. They are both
boundary
integrable quantum field theories with a quantum-group spin at the
boundary
which takes values, respectively, in standard or cyclic
representations of
the quantum group $SU(2)_q$.

This unification is powerful, and allows us to find new results for
both models. For the
anisotropic Kondo problem, we find exact expressions (in the presence
of a magnetic field) for all the
coefficients in the ``Anderson-Yuval''  perturbative expansion.
Our expressions hold  initially in the very anisotropic regime, but
we show how
to continue them beyond the Toulouse point all the way
to the isotropic point using an analog of dimensional regularization.
The analytic structure is transparent,
involving only simple poles which we
determine exactly, together with their residues. For the boundary
sine-Gordon model, which describes an impurity in a Luttinger liquid,
we find the non-equilibrium
conductance for all values of the Luttinger coupling.
This is an intricate
computation because the voltage operator and the boundary scattering
do not commute with each other.

\Date{9/95}

\newsec{Introduction}

One-dimensional quantum field theories with
gapless bulk excitations and boundary interactions
display a wide range of
interesting characteristics. They exhibit crossovers between
Fermi liquid and non-Fermi liquid behavior, they can be
sucessfully treated
by a variety of powerful and interesting
techniques, and they can be realized
experimentally.

The classic example of such a system consists of
electrons interacting with dilute impurities in
a metal, which can be described by the Kondo model.
This system is actually three-dimensional;
it can be described by a one-dimensional model because with dilute
enough impurities, the interesting physics occurs in $s$ waves
around each impurity, and one can restrict attention
to the radial coordinate.
This model has a variety of experimental realizations, and
has been the focus of much attention in the last 30 years
(see \refs{\AFL,\TW} and references within).
Of recent interest is the problem of an impurity in a
Luttinger liquid \KF. A Luttinger liquid (interacting electrons
in one dimension) may be realized in a one-dimensional wire
or by the edge of a fractional quantum Hall device \Wen.
A fractional quantum Hall device is made by putting an electron
gas trapped in two dimensions into a strong transverse magnetic
field. When the Hall conductivity
is locked to its plateau value,
the current flows only along the edges
of the device, and the system can be described effectively
by a one-dimensional theory.
Experiments have been done on the conductance through a point contact
(which is the impurity in the theory)
in one of these devices \exper, and they agree well with theory
\refs{\Moon,\FLS}.

The objects of our attention in this paper are one-dimensional
models with interaction on the boundary only.  We
concentrate on two such models,
the one-channel Kondo model
and the massless boundary sine-Gordon model.  The
problem of an impurity in a Luttinger liquid can be mapped
onto the latter.  Moreover,
when the bulk degrees of freedom are integrated out,
both describe problems in dissipative
quantum mechanics \Leggett: a particle
moving in a double well for Kondo (an infinite number of
wells for boundary sine-Gordon) with a dissipative environment.

In this work, we show that
the Kondo model and the boundary sine-Gordon model
can be treated in the same theoretical framework.
Both can be reformulated  as a free boson on the half-line
interacting with a spin on the boundary, where the spin
is in a representation of the ``quantum-group'' algebra $SU(2)_q$.
This algebra, as we will discuss below, is a one-parameter
deformation of the ordinary $SU(2)$ algebra.
In the Kondo model the spin is in a standard spin-$j$
representation, while for the boundary sine-Gordon model
the spin is in a ``cyclic'' representation,
a quantum-group representation
which has no analog in ordinary $SU(2)$. We will find
a simple relation between the partition functions of the two
models.
The relation is established through the use of the trace of the
quantum monodromy operator, an object generating the conserved
charges of the quantum KdV system \BLZ.  Having this relation, we
can relate quantities in one model to quantities in the other model.
For example, the perturbative coefficients of the partition function
of the spin$-1/2$ Kondo model are
expressed in terms of ordered integrals which are difficult to
evaluate.  This relation yields an expression for these
coefficients in terms of known coefficients of the boundary
sine-Gordon model \FLSjack.

The starting point of the theoretical analysis is a
one-dimensional quantum
theory at a fixed point of the renormalization group.
This means that the system has no mass scale, so there is no
gap in the spectrum. Such a model can be described by a
$1+1$-dimensional massless quantum field theory.
The issue of the
boundary conditions in these models is not a nuisance but
in fact can be  of crucial importance.
Basically, most of the physics which can
happen in the bulk can also happen on the boundary alone.
Studying boundary behavior is not only simpler mathematically,
but it can also be easier to observe experimentally.
There are very few experimental probes of one-dimensional
quantum systems, and the ones mentioned above are both boundary
effects.

A boundary fixed point is a point where the boundary condition
does not destroy the scale invariance of the bulk;
the methods of boundary conformal
field theory are applicable here. However, an
interacting boundary condition as in both the above systems will
introduce a scale to the problem, which we generically call
$T_B$ (in the Kondo problem this is often referred to as
the Kondo temperature $T_K$).
Although by definition bulk effects in these models do not
depend on this scale, boundary effects of a system at non-zero
temperature can now depend on the dimensionless parameter $T/T_B$.
Varying this parameter allows one to interpolate between different
boundary fixed points. For example, in the Kondo problem at
$T/T_B\to\infty$, there is a boundary fixed point where the impurity
decouples. As $T/T_B \to 0$, one approaches another boundary
fixed point where the electrons bind to the impurity. (The
properties of the low-temperature fixed point are far from obvious;
it took years  of effort to establish them.)
For the boundary-sine Gordon
model, the fixed points correspond to Dirichlet and
Neumann boundary conditions on a boson;
which is the high-temperature one and which
is the low-temperature one depends on the the boundary coupling.

The field theories we discuss have the special
property that they are integrable,
as are many one-dimensional theories.
As a result, we can do many calcluations exactly. In
this paper, we mainly discuss the partition function and free
energy. However, transport properties (which are experimentally
measurable) can also be computed exactly \refs{\FLS,\FLSbig}.
The methods we will describe
enable one to study these systems for all values of the coupling ---
near and far from the fixed points. Other methods generally
rely on perturbation theory around these fixed points.
Another advantage, for example, is that in the Luttinger problem
one can compute transport properties like the conductance even
out of equilibrium. Standard field theory techniques are not
applicable; the best one can do is use Kubo formula to calculate
the linear response near $V=0$.

The models discussed here can be treated as a boson
where the one-dimensional space is a half-line.
The only interactions take place on the boundary.
The bulk Hamiltonian takes the form
\eqn\genactbulk{H_0={1\over 4\pi g}\int_0^\infty d\sigma
\left[\Pi^2+(\partial_\sigma\phi)^2 \right]+{V\over
2\pi}\int_0^\infty
d\sigma\partial_\sigma\phi.}
The first model we discuss is the
boundary sine-Gordon model. The boundary hamiltonian is
\eqn\brdsg{H_{BSG}=2 v\cos\phi(0).}
In this case, the parameter $V$ of \genactbulk\
plays the role of a physical voltage, while $v$ is related to
the boundary scale $T_B$ in a manner to be discussed below.

The second model, the one-channel anisotropic Kondo problem of spin
$j/2$,
can also be expressed in this form using the well-known technique of
bosonization \Schotte.
We ignore the charge sector of the Hamiltonian, which does not
interact
with the spin and decouples from the problem. The total
Hamiltonian is then $H=H_0+H_j$, where the boundary interaction
$H_j$=
$\sum_{i=x,y,z} I_i J_i S_i$.  Here, $S_i$ is the impurity spin
on the boundary, $J_i$ are the fermion currents and $I_i$ are the
coupling constants. The problem is anisotropic when $I_z\ne I_x=I_y$.
In the bosonized language, the boundary Hamiltonian becomes
\eqn\genactbdry{H_j=\lambda\left(S_+e^{i\phi(0)}+
S_-e^{-i\phi(0)}\right)-{Vg\over 2}S_z.}
We have replaced the original parameters $I_x=I_y$ and $I_z$
with $g$ and $\lambda$;
$g$ parametrizes the anisotropy ($g=1$ is the isotropic case
and $g=1/2$ is called the Toulouse limit) while $\lambda\propto
|I_x|$.
The precise relation is not universal so we will not need it
here. The $J_zS_z$ term has been absorbed in a redefinition of $g$.
Traditionally in the Kondo problem, one takes the matrices
$S_i$ to act in the spin-$j/2$ representation of $SU(2)$
(in \genactbdry\ and in all what follows
we use the conventions that eigenvalues of $S_z$ are integer,
so they e.g.\ take values
$S_z=\pm 1$ in the spin $1/2$ representation). However,
for the problem to be integrable, one must instead take them to
act in the spin $j/2$ representation of
the quantum group $SU(2)_q$,
\eqn\qgrdef{[S_z,S_\pm]=\pm 2S_\pm,\quad
[S_+,S_-]={q^{S_z}-q^{-S_z}\over q-q^{-1}},}
where $q=e^{i\pi g}$ . In the isotropic case $q=-1$, this reduces
to the usual $SU(2)$ algebra. The distinction between
$SU(2)$ and $SU(2)_q$ is not important for
$j=1$ or $j=2$ at arbitrary $q$, because the spin
$1/2$ representation remains the Pauli matrices and the spin-$1$
representation is also the same up to a rescaling of $S_+$ and $S_-$.
In the following, the Kondo model
of spin-$j/2$ is defined to be the model with the $q$-deformed
algebra,
so it can be identified with the ``physical''
Kondo model only for $j=1,2$.
In the Kondo models, $\del_\sigma\phi$
is the $z$-component of the fermion current, so $V$ corresponds to
an external magnetic field.

We consider the system in imaginary time compactified
on a circle of length $1/T$ with $T$ the temperature.
Defining the partition function
via the trace ${\cal Z}_j=Tr\ e^{(H_0+H_j)/T}$,
we introduce $Z_j=(j+1){\cal Z}_j(\lambda)/{\cal Z}_j(0)$
and $Z_{BSG}={\cal Z}_{BSG}(\lambda)/{\cal Z}_{BSG}(0)$.
In the following,
we often use the variable $p$
defined as
\eqn\volta{ i{Vg\over T}=2\pi p.}
Strictly speaking,
the boson hamiltonians at non-zero temperature make sense only when
$p$ an integer. The consideration of real (physical)
voltage or magnetic field requires analytic continuation.

The dimension of the
vertex operators $e^{\pm i\phi}$ is $g$.
For example the two-point function on the boundary
is
\eqn\twopnt{ \langle e^{i\phi(0,\tau)}
e^{-i\phi(0,\tau')}\rangle =
\left|{\kappa\over\pi T}\sin{\pi T(\tau-\tau')}\right|^
{-{2g}},}
with $\kappa$ the frequency cutoff arising from the normal-ordering
of the operators $e^{\pm i\phi}$.
We will denote the
case $1/2<g<1$ as the repulsive regime and $0<g<1/2$ the
attractive regime; ``attractive'' and ``repulsive'' are
the corresponding type of fermion interactions when one fermionizes
this model into the Luttinger model \foot{In the quantum wire
problem (an impurity in a Luttinger liquid)
where one starts with electrons on a full line, the entire domain
$0<g<1$
corresponds to repulsive interactions between the physical electrons.
There is a rescaling
of the coupling when one maps the model on to the half-line \FLS.}.
The Toulouse limit
$g=1/2$ corresponds, of course, to free fermions.
For $g>1$ the vertex operators are irrelevant, and the model
is best approached by using a ``dual" picture \KF.

The paper is organized as follow.  In section 2 the attractive
regime is described.  Results are obtained to all orders
in perturbation theory
using Jack symmetric functions for the boundary
sine-Gordon model.  Then, making use of the monodromy matrix,
a relation is given between this latter model and the Kondo
models.  In the last part of this section, the thermodynamic
Bethe ansatz is used to provide non-perturbative results in
both cases.  In section 3, the repulsive regime is
explored.  There the perturbative coefficients of
the partition function diverge and
a regularization is needed, which usually is provided
by a high-frequency or short-distance
cut-off. We show, using the explicit
expressions discussed in section 2,
that these divergences can also be controlled by analytic
continuation
from the repulsive regime, an analog of dimensional regularization.
Coefficients for
the free energy can be obtained in this fashion all the way to $g=1$;
at particular values of $g$ there are poles, and we compute
the residues exactly.  These results are
in agreement with computations using the Bethe ansatz in the
repulsive regime, where the poles result in logarithmic terms
in the free energy.  In section 4, the relation between
models is extended to non-zero $V$. This gives the perturbative
coefficients in Kondo model as a function of magnetic field.
Moreover, it yields the previously-unknown
conductance for the boundary sine-Gordon model at
all values of $g$. Some final remarks are collected in the
conclusion.

\newsec{The attractive regime at zero voltage}

In this section we review earlier results for the anisotropic Kondo
problem at zero magnetic field and the boundary sine-Gordon (BSG)
model at zero voltage and with $g<1/2$.
There are three useful and complementary approaches, all of which
we will later extend to finite magnetic field (resp.\ finite voltage)
and to $g>1/2$.

We first discuss
how to expand the partition function in powers of the interaction
strength. For the Kondo problem, this was first considered  long
ago in \refs{\YA,\Schotte}, where the coefficients of this expansion
were expressed as multiple integrals. These integrals
are rather complicated, and until now had not been evaluated
explicitly except
in very special limits. The partition function in this
form is equivalent to that of a one-dimensional gas of positive
and negative charges with logarithmic interactions (equivalently of
a two-dimensional Coulomb gas on a circle). For
the boundary sine-Gordon model, the perturbative expansion
is formally very similar, but not identical. In that case,
the multi-dimensional integrals can be explicitly
evaluated, using recent results for symmetric polynomials \FLSjack.

The second approach uses integrability, albeit in a rather abstract
way. We define the trace of the
``quantum monodromy operator'' \BLZ, whose expectation value
gives the Kondo or BSG partition function, depending on which
representation is chosen \refs{\FSkon,\BLZnew}.
Using some properties of this
operator, we are able to relate the Kondo partition function to the
BSG
one. As a result we are able to evaluate the integrals in  the Kondo
expansion explicitly, using the
already-evaluated BSG ones \refs{\FSkon,\BLZnew}.

The third approach uses integrability in a more standard way.
We describe the model in terms of interacting quasiparticles and
their scattering matrices. The thermodynamic Bethe ansatz (TBA)
 can then be used to derive the free energy
and related quantities for the Kondo model \refs{\AFL,\TW}\
and for the BSG model \FSW. The direct relation between
the partition functions can be rederived, at least for values
of the coupling $g=1/t$, $t$ integer. The TBA approach
has the disadvantage that at non-zero temperature
the integral equations derived are not continuous in $g$
(although the final results of course are).
However, it has the
advantage that it allows transport properties
like the current and conductance \refs{\FLS,\FLSbig}\ and the
zero-temperature noise \FLSnoise\ to be computed for
the BSG model. (For Kondo,
only the zero-temperature magnetoresistance has been computed
\AFL.)
Some simple relations have also
been derived
relating transport properties to equilibrium
properties \FLSjack; we generalize these in sect.\ 4.

\subsec{Perturbative approach}

The partition functions $Z_j$ and $Z_{BSG}$ can be expanded
in powers of $\lambda$ and $v$ respectively.
The term of order $\lambda^{2n}$ or $v^{2n}$ involves a correlation
function of $n$  vertex operators $e^{i\phi}$ and $n$ vertex
operators $e^{-i\phi}$,
all living on the boundary.
These multi-point functions, evaluated in the
free-boson theory and by Wick's theorem, are reduced to a product
of two-point functions like \twopnt\ (see e.g.\ \ref\ISZ{C. Itzykson,
H. Saleur and
J. B. Zuber, {\it Conformal invariance and appplications to
statistical
mechanics}, World Scientific (1988).} for a review).
The problem then becomes formally equivalent to a two-dimensional
Coulomb gas with positive and negative charges restricted to
live on a one-dimensional circle. To calculate the partition
function,
we must integrate over the locations of the charges.
The integrand is then the scaled correlator

\eqn\corr{\eqalign{{\cal I}_{2n}(\{u_i\},\{u_i'\})&\equiv
\left({\kappa\over 2 \pi T}\right)^{2g}
\langle e^{i\phi(u_1/2\pi T)}\dots e^{i\phi(u_n/2\pi T)}
e^{-i\phi(u_1'/2\pi T)}\dots e^{-i\phi(u_n'/2\pi T)}\rangle\cr
&=\left|{\prod_{i<j}4\sin\left({u_i-u_j\over 2}\right)
\sin\left({u'_i-u'_j\over 2}\right)\over \prod_{i,j}
2\sin\left({u_i-u'_j\over 2}\right)
}\right|^{2g}. \cr}}
The difference between the Kondo model and the boundary sine-Gordon
(BSG) model lies in the limits of integration.
In the Kondo model,
each of the vertex operators comes with a spin operator;
the thermal  average of  monomials of vertex operators are
computed as in \corr, while one has to take the trace of the
corresponding monomial of spin operators in the representation
of interest. This puts various contraints on the order
of the charges.  For example, $S_+^2=S_-^2=0$ when the spin
is $1/2$, so only terms of the form $S_+S_-S_+S_-\dots$ survive
in the perturbative expansion, and consequently
charges alternate in sign on the circle. Thus in terms of
the renormalized parameter $x$, defined as
$$x \equiv  {\lambda\over T}
\left({2\pi T\over \kappa}\right)^{g},$$
the spin-$1/2$ Kondo partition function is \refs{\Schotte,\YA}
\eqn\expandit{Z_1(x)=2+\sum_{n=1}^\infty
x^{2n} Q_{2n},}
where
\eqn\Qdef{
Q_{2n}(p)=2\int_{0}^{2\pi}du_1\int_0^{u_1}
du'_1\int_0^{u'_1}du_2\ldots\int_0^{u_n}du_n'\
{\cal I}_{2n}(\{u_i\},\{u_i'\}).}
The effect of the charge ordering is seen in the limits of
integration.
Higher-spin partition functions have the same integrand, but
with the appropriate resctrictions on charge ordering.
For the boundary sine-Gordon model there is no boundary
spin, so one has unordered integrals:
\eqn\expandbsg{Z_{BSG}(x)=1+\sum_{n=1}^\infty
x_{BSG}^{2n} I_{2n},}
where
$$
x_{BSG}\equiv {v\over T}
\left({2\pi T\over \kappa}\right)^{g}
$$
and
\eqn\Idef{I_{2n}={1\over (n!)^2}
\int_0^{2\pi}du_1\ldots  \int_0^{2\pi} du'_n\
{\cal I}_{2n}(\{u_i\},\{u_i'\}).}
The lack of ordering makes no difference for $n=1$, so
$I_2=Q_2$, but the others are different \foot{
The $I_{2n}$ were denoted $Z_{2n}$ in \refs{\FLSjack,\FSkon};
we change notation here to avoid confusion with the higher-spin
partition functions.}.

The unordered integrals $I_n$ can be computed
exactly in terms of an $n$-dimensional series \FLSjack:
\eqn\newIII{I_{2n}=
{1\over [\Gamma(g)]^{2n}}\sum_{{\bf m}}
\prod_{i=1}^n \left({ \Gamma\left[m_i+g(n-i+1)\right]
\over \Gamma\left[m_i+g(n-i)+1\right]}\right)^2 ,}
where the sum is over all sets (Young tableaux)
${\bf m}=(m_1,m_2,\dots,m_{n})$, with integers $m_i$
obeying $m_1\ge m_2 \ge \dots m_{n}\ge 0$.
For $n=1$ the series can be summed,
giving $I_2=\Gamma(1-2g)/[\Gamma(1-g)]^2$.
Although this series looks quite imposing, it can be generated
by a simple recursion relation.  We introduce the
truncated sum $I_{2n}(\Lambda)$, which is defined as
the sum over ${\bf m}$ with the condition that all
$m_i\leq \Lambda$.
Then it is not difficult to show that
\eqn\recurrence{
I_{2n}(\Lambda)=I_{2n}(\Lambda-1)+\left({\Gamma(\Lambda+g n)\over
\Gamma(g) \Gamma (\Lambda+1+g(n-1))} \right)^2 I_{2(n-1)} (\Lambda).
}
These relations allow a precise determination of the partition
functions up to large orders in the perturbation expansion.

The boundary sine-Gordon model
can be placed in the same framework as the
anisotropic Kondo models. It is enough
to discuss the simplest case when $q$
is a root of unity ($g$ rational),
since by continuity, the results
we derive will hold for any $q$ of unit modulus.
Suppose therefore $q^k=\pm 1$ for some integer $k$,
and consider a cyclic  representation
of the quantum group $SU(2)_q$
(see \ref\Skl{E.K. Sklyanin, Funct. Anal. Appl. 16 (1983) 263; 17
(1983) 273.}). These representations, which
are labeled by an arbitrary complex parameter $\delta$,
have no highest- or lowest-weight state; the states are eigenstates
of $S_+$ or $S^-$ to the $t$th power. They have dimension $k$, with a
basis of states $|m>$ such that
\eqn\cycle{\eqalign{S_+|m>=&
{q^{(\delta-m)/2}-q^{-(\delta-m)/2}
\over q-q^{-1}}\ |m+1>\cr
S_-|m>=& {q^{(\delta+m)/2}-q^{-(\delta+m)/2}
\over q-q^{-1}}\ |m-1>\cr
S_z|m>=&m\ |m>,\cr}}
where states $|m>$ and $|m\hbox{ mod }k>$ are identified, and
the fundamental set is chosen to be $0,1,\ldots, (k-1)$.
To obtain the BSG model, we set $q^\delta=C$
 and let $C>>1$ and real (so $\delta$ is imaginary). Thus
\eqn\limitofgens{S_\pm |m>\approx C
{q^{\mp m/2}\over q-q^{-1}}|m\pm 1>,}
so in this limit the commutator of $S_\pm$ can be neglected and
the traces of all monomials become identical:
\eqn\limittr{Tr {\cal M}=k \left({
C q^{1/2}\over q-q^{-1}}\right)^{2n},}
where ${\cal M}$ is the product of
$n$ operators $S_+$ and $n$ operators
$S_-$ in any order.
Thus when evaluating $Z_\delta$ for $C$ large, all the possible
orderings of $S_+$ and $S_-$ within the trace have the same weight,
so
\eqn\useful{Z_{\delta}(x)
\approx k Z_{BSG}\left(C{q^{1/2}\over
q-q^{-1}}x\right),\qquad q^\delta=C>>1.}
This observation allows us, for example, to find the boundary $S$
matrix
of the BSG model \GZ, as we detail in the Appendix.
It will also enable us to derive many properties of the partition
function $Z_{BSG}$ in the subsequent sections.

\subsec{Quantum monodromy and fusion}

We introduce the quantum monodromy operators
associated with these models \BLZ\
\eqn\qmon{L_j(x)=\Pi_j\left\{e^{i\pi  PS_z}{\cal P}\exp\left[
q^{-1/2}x \int_{0}^{1/T}d\tau \left(e^{-2i\phi_L(\tau)}q^{S_z/2}S_+ +
e^{2i\phi_L(\sigma)}q^{-S_z/2}S_-\right)\right]\right\},}
where $\Pi_j$ indicates that the matrices $S_j$ are in the
spin-$j/2$ representation, ${\cal P}$ indicates path ordering,
and the exponentials are normal-ordered.
In this formula, $P$ is the momentum operator appearing
in the mode expansion of the left-moving field $\phi_L$
$$\phi_L(\tau)=Q+ 2\pi TP\tau+i\sum_{n\neq 0}
{a_{n}\over n} e^{-2i\pi n T\tau}.$$
Here the normalizations are $[Q,P]=ig$,
$[a_n,a_m]={ng\over 2}\delta_{n+m}$.
Observe that $L_j$ is an operator acting both on
the spin degrees of freedom
and on the ``free-boson''
degrees of freedom. By expanding $L_j$ in powers of
$x$ and noting that with
Neumann boundary conditions $2\phi_L(0,\tau)=\phi(0,\tau)$,
one finds that the partition functions are equal to the
eigenvalues of the quantum transfer matrices acting
on momentum eigenstates $P|p\rangle ={p}|p\rangle$,
\eqn\identt{Z_j(x,p)=<p| \tr\  e^{i\pi PS_z}L_j(x) |p>,}
where the trace is computed over the spin degrees of freedom.
At zero voltage, $p=0$; we define $Z_j(x)\equiv Z_j(x,p=0)$.

As observed in \BLZ, the $L_j$ satisfy the
Yang-Baxter equation. Using the fusion
of quantum transfer matrices,
one can prove the identities
\eqn\fusion{\eqalign{&Z_j(q^{1/2}x)Z_j(q^{-1/2}x)=1+Z_{j-1}(x)
Z_{j+1}(x)\cr
&Z_{1}(q^{(j+1)/2}x)Z_j(x)=Z_{j+1}(q^{1/2}x)+Z_{j-1}(q^{-1/2}x)\cr
&Z_1(q^{(\delta+1)/2}x)Z_\delta(x)=
Z_{\delta+1}(q^{1/2}x)+Z_{\delta-1}(q^{-1/2}x).\cr}}
The first was discussed in \BLZ; the second can by proven by
using the first and by induction. The last follows using the
same technique as in \ref\KR{A.N. Kirillov and N. Yu. Reshetikhin,
J. Phys. A20 (1987) 1565, 1587.}, together with the fusion rules
for cyclic and standard representations
\ref\DJMM{E. Date, M. Jimbo, K. Miki and T. Miwa, Comm. Math.
Phys. 137 (1991) 133.}

Using these relations together with \useful, we can express the
boundary
sine-Gordon model partition function in terms of the Kondo partition
function. We have, from \fusion,
$$
Z_1(C q^{1/2}x)Z_{BSG}\left({C q^{1/2}\over q-q^{-1}}x
\right)
=Z_{BSG}\left({Cq^{3/2}\over q-q^{-1}}x\right)
+Z_{BSG}\left({Cq^{-1/2}\over q-q^{-1}}x\right)
$$
from which it follows that
\refs{\FSkon,\BLZnew}
\eqn\fusiongoodie{Z_1[(q-q^{-1})x]={Z_{BSG}(qx)
+Z_{BSG}(q^{-1}x)\over Z_{BSG}(x)}.}
Inserting the perturbative expansions into \fusiongoodie\ gives
the $Q_{2n}$ in terms of the already-known $I_{2n}$, thus completing
the derivation of the perturbative partition function for $g<1/2$.

\subsec{The thermodynamic Bethe ansatz}

The fusion relations discussed in the previous subsection are
one of the many consequences of integrability \ref\Bax{R. J. Baxter,
{\it Exactly solved models in statistical mechanics}, Academic Press
(1982).}.
The standard way of approaching the problem is to use the
Bethe ansatz. Here, one derives integral equations which determine
a set of functions $\epsilon_j(\theta)$, where $\theta$ is a
rapidity (the logarithm of the energy of an individual particle).
The $\epsilon_j(\theta)$ can be thought of as
the energy of an interacting quasiparticle, in the
sense that the energy of the entire system shifts by
$T\epsilon_j(\theta)$ when a particle of rapidity $\theta$
is added to the system. Moreover, the distribution function
is given by $1/(1+\exp(\epsilon_j))$. Many physical
quantities can be expressed in terms of these functions.
Since this approach has been discussed in detail in many places,
we start with the integral equations and discuss their consequences.
For technical reasons, we consider the case $g={1/ t}$,
where $t$ is an integer. The integral equations for both Kondo
and BSG are \refs{\AFL,\TW,\FSW}
\eqn\summeqatt{
\epsilon_j=\sum_k N_{jk}\int_{-\infty}^{\infty} {d\theta'\over 2\pi}
{t-1\over \cosh[(t-1)(\theta-\theta')]}
\ln\left(1+e^{\epsilon_k(\theta')}\right)}
where the ``incidence matrix'' $N_{jk}$ is defined by the diagram
\bigskip
\noindent
\centerline{
\hbox{\rlap{\raise28pt\hbox{$\hskip5.5cm\bigcirc\hskip.25cm +$}}
\rlap{\lower27pt\hbox{$\hskip5.4cm\bigcirc\hskip.3cm -$}}
\rlap{\raise15pt\hbox{$\hskip5.1cm\Big/$}}
\rlap{\lower14pt\hbox{$\hskip5.0cm\Big\backslash$}}
\rlap{\raise15pt\hbox{$1\hskip1cm 2\hskip1.3cm k\hskip.8cm t-3$}}
$\bigcirc$------$\bigcirc$-- -- --
--$\bigcirc$-- -- --$\bigcirc$------$\bigcirc$\hskip.5cm $t-2$ }}

\bigskip

\noindent
where $N_{jk}=1$ if the nodes $j$ and $k$ are connected, and
$N_{jk}=0$
if not.
The solution of these integral equations is fixed uniquely by
demanding the asymptotic form
\eqn\asym{ \epsilon_j\approx 2\sin {j\pi\over 2(t-1)}e^{\theta},
\qquad \epsilon_\pm \approx e^\theta
\quad \hbox{as }\theta\to \infty.}
This asymptotic form is just the energy of the individual particle
over $T$;
the interactions become negligible in the large-energy limit
(equivalent
to sending $\lambda$ and $v$ to zero in the Hamiltonian).

The free energies in this regime for Kondo \TW\
(with spin less than $t/2$) and BSG \FSW\ can
be written in  the form
\eqn\fkon{F_{j}=T_B{\sin [j\pi/2(t-1)]\over \cos[\pi/2(t-1)]}-
T\int {d\theta\over 2\pi}
{t-1\over \cosh[(t-1)(\theta-\ln T_B/T)]}
\ln\left(1+e^{\epsilon_j}\right),}
for $j=1\dots t-2$ together with $F_{t-1}=2 F_{BSG}$ and
\eqn\fbsg{F_{BSG}=
{T_B\over 2\cos[\pi/2(t-1)]}-T\int {d\theta\over 2\pi}
{t-1\over \cosh[(t-1)(\theta-\ln T_B/T)]}
\ln\left(1+e^{\epsilon_{t-1}}\right),}
where $\epsilon_{t-1}\equiv \epsilon_+=\epsilon_-$.

It has been shown that the equations \summeqatt\ require that
$\epsilon_j(\theta)=\epsilon_j(\theta + i2\pi t/(t-1))$
\periodtba. This means that the integrals in \fkon\ and \fbsg\
can be expanded as a power series in $(T_B/ T)^{2(t-1)/t}$,
so we see that the bare couplings $\lambda$ and $v$ and
the renormalized coupling $x$ are proportional to $T_B^{(t-1)/t}$.
In fact, for BSG,
the exact constant was determined in \FLSbig, and is
\eqn\bsgnorm{
x_{BSG}\equiv  {v\over T}
\left({2\pi T\over \kappa}\right)^{g}=\Gamma(g)
\left({T_B\over T}
{\Gamma({1\over 2(1-g)})\over 2\sqrt{\pi}
\Gamma({g\over 2(1-g)})}\right)^{1-g},}
for any value of $g$, not just $g=1/t$.
At fixed $T_B$, the Kondo bare coupling $\lambda$
is related to the bare BSG coupling
via a constant to be determined at the end
of this section: $\lambda\equiv \xi v$.
This constant $\xi$ is independent
of the impurity spin considered, as observed in \TW.
With this relation of $x$ and $T_B$, we see that the second
term in \fkon\ or \fbsg\ is an analytic power series
in $x$, like the perturbative partition functions in
sec.\ 2.1.

We must take care in relating these non-perturbative free energies
to the perturbative partition functions discussed before. The TBA
deals with
excitations over the vacuum; by convention, the ground state (no
particles)
is assigned a vanishing energy and entropy. Therefore,
one expects $F_j$ and
$F_{BSG}$ to equal to the perturbative partition functions
defined previously up to a constant shift (which on dimensional
grounds must be proportional to $T_B$)
and a term proportional to $T$.
Neither of these change e.g.\ the specific heat.
This ambiguity is fixed
by studying the behavior at $T_B=0$ and by studying the analyticity
properties. From the TBA equations, it is simple
to derive from \summeqatt\ that as $\theta\to -\infty$, the functions
$\epsilon_j(\theta)$ go to a constant, which is
\eqn\Tzero{ e^{\epsilon_j(-\infty)}= (j+1)^2 -1, \qquad
e^{\epsilon_\pm(-\infty)}= t-1.}
 Plugging this into
\fkon\ gives
$$F_j(T_B=0)=-T\ln(1+j), \qquad F_{BSG}(T_B=0)=
-{T\over 2}\ln t.$$
This fixes the piece proportional to $T$. The piece proportional
to $T_B$ is fixed by noticing that because
the perturbative partition function is analytic
in $x\propto T_B^{1-g}$, the term proportional to $T_B$
cannot appear here.
Thus the relation between the
perturbative partition functions and the TBA free energies is
\eqn\tbapert{\eqalign{F_j&=-T\ln Z_j
+T_B{\sin [j\pi/2(t-1)]\over \cos[\pi/2(t-1)]} \cr
F_{BSG}&=-T\ln Z_{BSG}+{T_B\over 2\cos[\pi/2(t-1)]}-{T\over 2} \ln
t\cr}}
The role of the shift is simple: it precisely cancels the large
$T_B/T$
behavior of the partition functions. As is easily seen by
substituting
the asymptotic form \asym\ into \fkon\ and \fbsg,
the free energies $F_j/T$ and $F_{BSG}/T$ determined by the TBA
go to zero as $T/T_B\to 0$.  Meanwhile it was shown in \FLSjack\
that $-\ln Z_{BSG} \propto T_B/T$ in this limit, and the relations
\fusiongoodie\ and \fusion\ indicate that $-\ln Z_j$ grows as well.
In fact from \tbapert\ it follows that
$$
\eqalign{Z_j\approx &\exp \left({T_B \sin [j\pi/2(t-1)]\over
T\cos[\pi/2(t-1)]}\right)\cr
Z_{BSG}\approx & {1\over\sqrt{t}}\exp \left({T_B\over
2T\cos[\pi/2(t-1)]}\right).\cr}
$$
Thus we see that although the perturbative partition functions grow
exponentially for large $x$, the series expressions are still
convergent (they actually have an infinite radius of convergence).

Since we have related the TBA results
to the perturbative ones, we can combine the
the fusion relations \fusion\ with the TBA
equations \summeqatt\ to give much more information.
For example, the relation \fkon\ gives the Kondo partition functions
$Z_j$ only for $j=1\dots t-2$, but the remainder can be
generated from \fusion.
The relation \tbapert\ allows us to write the
perturbative partition functions in terms
of the $\epsilon_j$ very simply.
Denoting convolution by
$$A*B(\alpha)= \int_{-\infty}^{\infty}
{d\theta\over 2\pi} A(\alpha-\theta)B(\theta),$$
one has
\eqn\zfrels{\eqalign{
\ln Z_j(x)=&  s_{t-1}*\ln (1+e^{\epsilon_{j}}) (\alpha)
\qquad j=1\dots t-2\cr
\ln Z_{BSG}(\xi x)=&-{1\over 2}\ln t+
s_{t-1}*\ln (1+e^{\epsilon_{t-1}}) (\alpha)\cr
\ln Z_{t-1}(x) &=2s_{t-1}*\ln (1+e^{\epsilon_{t-1}}) (\alpha),\cr
}}
where $s_{a} = a/\cosh(a\theta)$ and $\alpha=\ln T_B/T$.
These relations also give immediately
\eqn\magic{Z_{t-1}(x)=t\ Z^2_{BSG}(\xi x).}
In the following, we often switch from the $x$ variable
to $\alpha$, keeping these relations in mind.
Introduce as usual
\eqn\deff{Y_j(\alpha)=e^{\epsilon_j(\alpha)}.}
The $Y_j$ are analytic functions of
$x^2 \propto \exp[2t\alpha/(t-1)]$ \periodtba.
{}From the TBA equation for $\epsilon_1$ one finds then
$Y_1(\alpha)=Z_2(x)$
and from the TBA relation for $\epsilon_{t-1}$,
$Y_{t-1}(\alpha)=Z_{t-2}(x).$
The TBA relations for the other nodes then imply that
\eqn\magicii{Y_j(\alpha)=Z_{j+1}(x)Z_{j-1}(x),
\qquad j=1,\ldots, t-2.}
This simple relation between
the TBA and the perturbative partition function
gives exact perturbative expressions for all the $\epsilon_j$.
This is consistent with the relation \magic,
since the original integral equations \summeqatt\ along
with \zfrels\ give
$Y_{t-2}= tZ_{t-3} Z^2_{BSG}$.

We can
in fact rederive the results of sect.\ 2.2
at $q=e^{i\pi/t}$ by converting the
TBA equations \summeqatt\ into functional equations
in the complex $\alpha$ plane.
Using the identity
\eqn\sident{s_{a}(\theta +i{\pi\over 2a}) +
s_{a}(\theta -i{\pi\over 2a})
=2 \pi \delta(\theta).}
with \zfrels\ gives
\eqn\zy{Z_j(q^{1/2}x)Z_j(q^{-1/2}x)=1+Y_j(\alpha).}
where $j=1\dots t-2$.
Plugging the relation \magicii\ into \zy\ recovers
the fusion relation \fusion
\eqn\trans{Z_j(q^{1/2}x)Z_j(q^{-1/2}x)
=1+Z_{j-1}(x)Z_{j+1}(x)}
from \BLZ. Once written in terms of the variable $q$, \trans\ holds
for $q$ generic, as discussed in sect.\ 2.2.
Observe that such a direct proof of
\trans\ establishes conversely
the fact that the relation between $T_B/T$ and
$x$ is independent of spin.

When $q=e^{i\pi/t}$, $t$ an integer, these fusion relations
close \foot{A different closure happens in the minimal
models of conformal field theory \BLZ, where the end nodes
$t-2$, $+$ and $-$ are
removed from the incidence diagram.}.
We have from \trans
$$
Z_{t-1}(q^{1/2}x)Z_{t-1}(q^{-1/2}x)=
1+Z_{t-2}(x)Z_{t}(x)
$$
while using \sident\ in \zfrels\ gives
\eqn\canyoubelieveit{Z_{BSG}(\xi q^{1/2}x)Z_{BSG}(\xi q^{-1/2}x)=
{1\over t} (1+Y_{t-1}(\alpha)).}
Thanks to the identification $Y_{t-1}=Z_{t-2}$
these two relations are compatible with \magic\ if and only if
\eqn\moremagic{Z_{t}(x)=Z_{t-2}(x)+2.}
The latter relation follows from the quantum group representation.
Indeed, when $q$ is a $t^{th}$ root of unity,
the representation of spin $t$ is reducible
because $S_{\pm}^t=0$, and looks
schematically as in figure 1.
\fig{A schematic representation of a cyclic representation of
$SU(2)_q$
when $q$ is a $t^{th}$ root of unity. Up and down arrows represent
the action of raising and lowering generators.
}{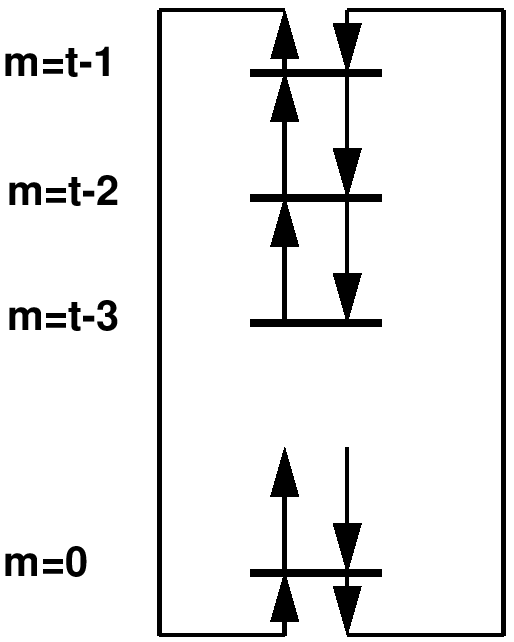}{6cm}{1}
\figlabel\tabb
We see that the states with values $S_z=\pm t$
do not contribute to the trace of any monomial
in $S_+S_-$ of non vanishing
order. Hence in the perturbative expansion, all terms
for spin $t-2$ and spin $t$ are equal, except the term of
order zero that simply counts the number of states. This
term  differs by two in the two representations, and
\moremagic\ follows.

We can also rederive the relation \fusiongoodie\ between the
Kondo and BSG partition functions without using
cyclic representations by
using the fusion relation from \fusion:
$$
Z_{1}(ix)Z_{t-1}(x)=Z_{t-2}(q^{-1/2}x)+
Z_{t}(q^{1/2}x).$$
Using \moremagic, we rewrite the right hand side as
$$
1+Z_{t-2}(q^{1/2}x) +1+Z_{t-2}(q^{-1/2}x),
$$
which using $Y_{t-1}=Z_{t-2}$ and \canyoubelieveit\ this is in turn
$$
tZ_{BSG}(\xi x)[Z_{BSG}(\xi qx)+Z_{BSG}(\xi q^{-1}x)]
$$
from which, using \magic, it follows that
\eqn\nice{Z_1[(q-q^{-1})x]={Z_{BSG}(qx)+Z_{BSG}(q^{-1}x)\over
Z_{BSG}(x)},}
together with the fact that $\xi={i\over (q-q^{-1})}$.
This value for $\xi$
is obtained simply by matching the first order in the perturbation
theory,
since we know that $I_2=Q_2$.

\newsec{The repulsive regime at zero voltage}

\subsec{Poles and log terms in the perturbative expansion}

The previous section concerned the attractive regime $g<1/2$.
This, for example, is the regime of greatest interest in
dissipative quantum mechanics, where the particle exhibits
oscillatory behavior. The filling fractions $\nu=1/(2n+1)$
where the edge modes in the fractional quantum Hall effect
are described by the BSG model also lie in this regime.
However, the original isotropic Kondo
model is at $g=1$, and we will see that there is a great
deal of interesting behavior in the repulsive regime $1/2<g<1$.
We will address this  regime largely  by exploiting some simple
analyticity
properties. In particular, we will show how to obtain
an analytic expression for
the coefficients all the way to $g=1$.

We will show in this section that if
we define the expansion of the free energy of the spin-$1/2$
Kondo model as
$$-T\ln Z_1=T\sum_{n=0}^\infty f_{2n} x^{2n}=
-T\ln (2+ \sum_{n=1}^\infty Q_{2n}x^{2n}),$$
the coefficient
$f_{2n}$ has a simple pole at $g=1-1/(2n)$, with residue
$r_{2n}=-1/( 2\pi n^2)$. The half-integer-spin Kondo and BSG
free energy expansions
have poles in the same places.
Moreover, when this
divergence is regulated properly, we find a term
$-2n r_{2n} T_B \log (T_B/T)$
in the free energy for $g=1-1/(2n)$. This term yields, for example,
a term linear in $T_B/T$ in the specific heat, indicating
that these values of $g$ (which include the Toulouse limit $g=1/2$)
are pathological in some respects \foot{However, notice that now
a term $T^{2n(1-g)}$ appears in the specific-heat
expansion at all $g$; the log term is required to make this true
at $g=1-1/2n$. Thus in this sense the log terms make
are not pathological but instead make
the values $g=1-1/(2n)$ {\it more} like other values of $g$.}.

The first thing to notice is that the integrals \Qdef\ and
\Idef, which define the perturbative coefficients $I_{2n}$ and
$Q_{2n}$,
diverge at short distances when $g\ge 1/2$; correspondingly,
the series expansion \newIII\ diverges for $g\ge 1/2$.
There are a variety of ways to regulate the integrals.
In a numerical approach, this would be done using a cut-off.
However, the most natural approach here is
analytic continuation. This approach, which is
very analogous to dimensional
regularization \ref\Zinn{J. Zinn-Justin,
{\it Quantum field theory and critical phenomena},
Oxford (1989).},
means we {\it define} the regularized integrals as the
analytic continuation of their values for $g<1/2$.
This continuation
is illustrated by examing the first coefficient, $f_2=-I_{2}/2$,
which we saw in sect.\ 2.1 is given by
$I_2=\Gamma(1-2g)/[\Gamma(1-g)]^2$.
At $g=1/2$, $f_2$ has a simple pole. There
are no branch points anywhere, and since it
is finite for all other $g\le 1$, the analytic
continuation is perfectly well-defined.
Implicit in the following is the assumption that the
regularization done for the Bethe ansatz
(the cutoff of the Fermi sea for Kondo \TW)
gives the same results as this analytic
continuation from $g<1/2$.
This assumption is certainly physically
obvious, since by defining renormalized parameters one
removes all cutoff dependence from the Bethe ansatz.
Moreover, in the BSG model one starts directly from the regulated
theory with no cutoff dependence \GZ.

We now show that the large-$T_B/T$ behavior of the partition function
requires that $f^{(BSG)}_{2n}$ and $f_{2n}$ have simple poles
at $g=1-1/(2n)$ for all $n$. As discussed in
sect.\ 2.3, in this limit $-T\ln Z_{BSG}$ behaves like
$$-T_B {1\over 2 \cos\left(\pi g\over 2(1-g)\right)},$$
while $-T\ln Z_1$ goes as
$$-T_B \tan\left(\pi g\over 2(1-g)\right).$$
Our analyticity assumption implies that these hold for
$g\ge 1/2$ as well.
Notice that these expressions have a simple pole as $g\to 1-1/(2n)$.
In the TBA free energy this term is subtracted
off, as seen in \tbapert. Because the TBA is finite, this
divergence therefore is matched by one in the perturbative
expansion.
Since $x\propto (T_B/T)^{1-g}$,
the terms $f^{(BSG)}_{2n}x^{2n}$ and $f_{2n} x^{2n}$
are proportional
to $T_B$ when $g=1-1/(2n)$. Therefore, there must be a simple
pole in $f_{2n}x^{2n}$ at $g=1-1/(2n)$,
with residue $r_{2n}\equiv T_B/(2\pi n^2 T)$.
The pole in $f^{(BSG)}_{2n}x^{2n}$ has residue
$(-1)^{n+1} r_{2n}/2$.
By the same argument, the free energy
coefficients in the spin-$j/2$
Kondo model when $j$ is odd
each have a single pole at $g=1-1/(2n)$  with residues $r_{2n}$.
The free energy coefficients for the
integer-spin Kondo model have no pole at these values.

We can see these poles explicitly by studying
the series expansion \newIII\ for the $I_{2n}$ in the
boundary sine-Gordon model. Initially the series looks useless,
because it diverges for $g\ge 1/2$, and we only
know how to resum it for $I_2$.
However, a first interesting observation is
that the series expressions for the $f^{(BSG)}_{2n}$
converge even where those for the individual $I_{2n}$ do not.
This is because some of the divergences in the Coulomb
integrals are cancelled when taking  the connected part.
More precisely,
we define the truncated series $I_{2n}(\Lambda)$ as the expression
\newIII\ with all $m_i\le \Lambda$ and
$$f^{(BSG)}_{4}(\Lambda)\equiv \left[{(I_2(\Lambda))^2\over 2}
-I_4(\Lambda)\right].$$
Then, using the previously obtained recurrence relation
\recurrence,  one finds that
\eqn\diver{
f^{(BSG)}_{4}(\Lambda)-f^{(BSG)}_{4}(\Lambda-1)\simeq \left(
{2g (1-2g)+1 \over 2 (1-2g)}
\right)\Gamma(g)^{-4} \Lambda^{4g-4}
}
for $\Lambda$ large. This expression converges as $\Lambda\to\infty$
for $g<3/4$.
Moreover, the pole at $g=3/4$ is clearly identified,
and its residue can easily be computed,
because the divergence is  proportional to that
of the zeta function. One confirms the earlier result
that near $g=3/4$
$$f^{(BSG)}_{4}x^4\approx -{1\over 16\pi(g-3/4)}
\left({T_B\over T}\right)^{4(1-g)},$$
where we used the relation \bsgnorm\ to relate $x$ and $T_B$.

Since \diver\ tells us explicitly how the series diverges, the
continuation around the pole
can be constructed by adding and subtracting
a zeta function. More precisely, we define the continuation to be
\eqn\regu{f^{(BSG)}_4(\infty)=f_4^{(BSG) reg}(\infty)+
{2g (1-2g)+1 \over 2 (1-2g) \Gamma(g)^{4}}\zeta(4-4g).
}
where
$$
f^{(BSG) reg}_{4}(\Lambda)\equiv f^{(BSG)}_{4}(\Lambda)-
{2g (1-2g)+1 \over 2 (1-2g) \Gamma(g)^{4}}
\sum_{n=1}^\Lambda n^{4g-4}.
$$
The ``regular'' part
$f^{(BSG) reg}_{4}$ is $f^{(BSG)}_4$ with the diverging part of the
sum
subtracted off. This series
converges when $\Lambda\rightarrow \infty$ for $g<1$.
One can extend this result past $g=1$ in the same manner.

It should be possible to find all the $f_{2n}$ for $g<1$
in this manner.
One first writes the higher $f_{2n}$'s in terms
of the $I_{2n}$ using the relation
\eqn\cumulant{
f^{(BSG)}_{2n}=\sum_{\bf m} {(-1)^{l({\bf m})-1}
(l({\bf m})-1)! \over \prod_j \lambda_j! } I_{2(n)}
}
where ${\bf m}=\{m_1,m_2,\dots m_{l({\bf m})}\}$ is
a partition of an integer so that $\sum m_i =n$,
$I_{2(n)}\equiv
I_{2m_1} I_{2 m_2} \cdots $
and $\lambda_j$ is the multiplicity of the integer $j$
in ${\bf m}$. We have checked that
$f_6^{(BSG)}(\Lambda)-f_6^{(BSG)}(\Lambda-1)\rightarrow
C_6(g)\Lambda^{6g-6}$
and
$f_8^{(BSG)}(\Lambda)-f_8^{(BSG)}(\Lambda-1)\rightarrow
C_8(g) \Lambda^{8g-8}$ when $\Lambda$ is large, with
$C_6(g), C_8(g)$ known expressions. Thus poles
in $f_{2n}^{(BSG)}$ appear at $g=1-1/2n$
for $n=1,2,3,4$ with the appropriate residue. We can then
apply the same zeta-function method and regularize the
sums to go all the way to $g=1$.

We have checked that the numerical
values agree very well with the Bethe ansatz results.
This takes some effort because finding the numbers from
the Bethe ansatz requires that we numerically solve the integral
equations, and then numerically fit the results to a power series.
Moreover, at $g=1$, we have $Z_{BSG}(x)=1$,
We plot the results for $f_4^{(BSG)}$ and $f_6^{(BSG)}$ in figs.\
2 and 3.
\fig{The free-energy coefficient $f^{(BSG)}_4$ as a function of
$g$.
The pole is at $g=3/4$.
}{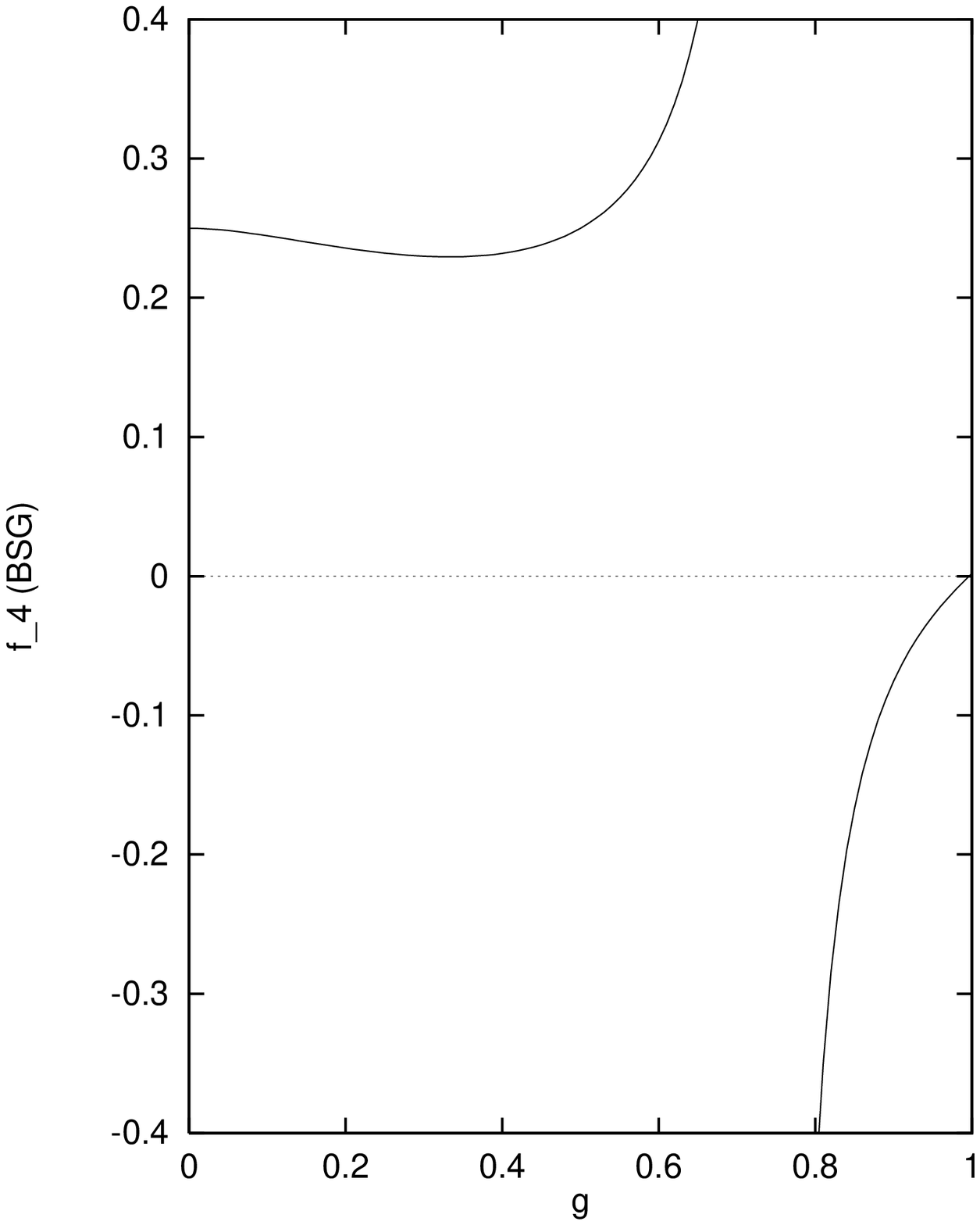}{6cm}{2}
\figlabel\tabb
\fig{The free-energy coefficient $f^{(BSG)}_6$ as a function of
$g$. The pole is at $g=5/6$.
}{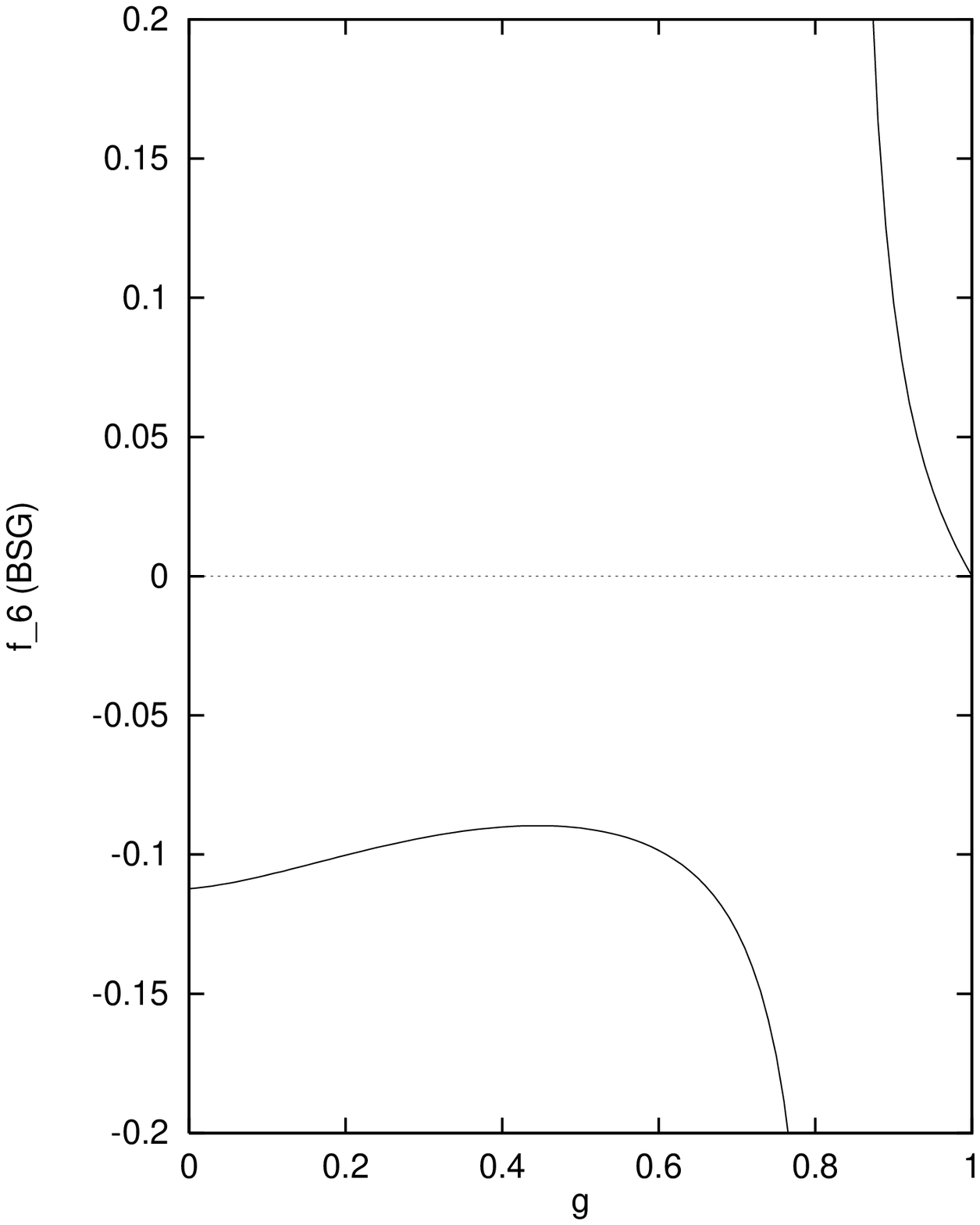}{6cm}{3}
\figlabel\tabb
We clearly see the pole in the data, and that
$f^{(BSG)}_4$ and $f^{(BSG)}_6$
do indeed go to zero as $g\to 1$.
Another interesting
consequence is that it allows a very simple approximation formula
for the $f_{2n}$ or $f^{(BSG)}_{2n}$ for $g$ near $1$.
We approximate the function
by its pole plus a constant piece. For example, if
we define the constant piece by requiring that
$f^{(BSG)}_{2n}(g=1)=0$,
we have
$$f^{(BSG)}_{2n}(g)\approx {\Gamma(n-1/2)\over 2\sqrt{\pi} n^2
\Gamma(n)\Gamma(1-1/2n)^{2n}}
\left[2n+{1\over 1-g-(1/2n)}\right].$$

The presence of these  poles  has interesting physical consequences.
They indicate that the free energy
at $g=1-1/(2n)$ cannot be expanded as a power series, but has  an
additional  logarithmic term.
The TBA free energy does not have a divergence, because
the pole is subtracted off, as in \tbapert.
However, there is a leftover piece:
$$\eqalign{lim_{g\to 1-1/(2n)}
\left[T_B \tan\left({\pi g\over 2(1-g)}\right) \right.&+ \left.
{Tr_{2n}\over g-1+1/(2n)} \left({T_B\over T}\right)^{2n(1-g)-1}
\right]\cr
&=-2nT r_{2n} \ln \left({T_B\over T}\right)+\ldots .\cr}$$
Thus the free energy
contains a logarithmic correction, arising
from proper regularization of the divergence.
Such terms are not unusual; for example they occur in the bulk free
energy of the 2D Ising model and its
(even) multicritical generalizations
\ref\ABF{G. E. Andrews, R. J. Baxter and P. J. Forrester,
J. Stat. Phys. 35 (1984) 193.}. As we will see in the
next subsection, the existence of the log term also follows from the
detailed TBA analysis.
The fact that the
free energy defined by analytic continuation has a simple pole
at $r_{2n}$ does not mean that the physical free
energy -- obtained with a cut-off regularization -- diverges,
but rather indicates that it has a logarithmic dependence on the
cut-off
at that point \Zinn.

These results can be checked in the   Toulouse limit $g=1/2$,
where the model is equivalent to a free fermion in a boundary
magnetic field. As discussed in \TW\ for example, the free energy
for spin-$1/2$ Kondo is
$$\eqalign{F_{2}&=-2T\int_{-\infty}^{\infty} {d\theta\over 2\pi}
{1\over
\cosh(\theta-\ln T_B/T)} \ln (1+e^{-e^\theta})\cr
&=2T\ln\Gamma\left({T_B\over\pi T}\right)-
2T\ln\Gamma\left({T_B\over 2 \pi T}\right)+
{T_B\over \pi }\left[1-2\ln 2 -\ln {T_B\over 2 \pi T}\right]
-T\ln 2.\cr}$$
Using the gamma function identity
$\Gamma(2a)=\Gamma(a)\Gamma(a+1/2) 2^{2a-1}/\sqrt{\pi}$
and the expansion
$$\ln\Gamma(a+1/2) = \ln\Gamma(1/2)
+ \sum_{n=1}^\infty \psi^{(n-1)} (1/2) {a^n\over n!},$$
where $\psi^{(m)}(x)$ is the $m$th derivative of
the digamma function $\psi(x)\equiv\Gamma'(x)/\Gamma(x)$, one finds
\eqn\ghalf{F_2=-T\ln 2 +{T_B\over \pi }(1-\ln{T_B\over 2\pi T}) +2T
\sum_{n=1}^\infty \psi^{(n-1)} (1/2){1\over n!} \left({T_B\over 2\pi
T}\right)^n.}
Thus we see explicitly the log term at $g=1/2$, with coefficient
$r_2=1/2\pi$ as derived above. Moreover, we see that all the
$f_{2n}$ for $n>1$ are finite at $g=1/2$.
This means, for example, that
there is a double pole in $Q_4$ at $g=1/2$ in order for
$f_4=-Q_4/2+Q_2^2/8$ to remain finite. In fact, this means that there
is an $n$th order pole in $Q_{2n}$ at $g=1/2$ and that they are
analytic in the neigborhood (and in fact all the way to $g=3/4$).
Moreover, at $g=1/2$, $F_{BSG}=F_1/2$, and we have checked
numerically that the values for $f^{(BSG)}_{2n}$
from \ghalf\ are obtained by taking the limit of the series
expression as $g\to 1/2$ \newIII.

A final comment is in order. At the isotropic point $g=1$, the
BSG model is trivial with these boundary conditions, so
$Z_{BSG}(x)=1$
and $I_{2n}=0$. Notice that this follows easily from the
relation \fusiongoodie.
However, the Kondo problem is not trivial at $g=1$
(this is the value of most physical interest), but the power series
expression (proportional to $T_B^{1-g}$) obviously requires
modification.
The fact that the exponent is vanishing is an obvious hint that
there are log terms at every order, and indeed this is seen in
the TBA solution \refs{\AFL,\TW}.
Notice that the shift between the TBA and
the power series has an essential singularity as $g\to 1$, so
if subtracted appropriately from the power series
as $g\to 1$, the result may be finite and give the series with log
terms at $g=1$. We have not yet succeeded in carrying out this
analysis.

\subsec{The TBA in the repulsive regime}

The TBA equations for the Kondo problem in the repulsive
regime were derived in \TW. For technical simplicity, we
consider only $g=1-1/s$, $s\ge 2$ an integer.
The equations are very similar to those in the attractive
regime:
\eqn\summeqrep{\epsilon_j=\delta_{j1}e^{-\theta}
-\sum_k N_{jk}\int {d\theta'\over 2\pi}
{1\over\cosh(\theta-\theta')}
\ln\left(1+e^{-\epsilon_{k}(\theta')}\right),}
where the incidence matrix $N_{jk}$ is as in sec.\ 2.3 with
$t$ replaced by $s$.
The Kondo free energy is
\eqn\sumfimp{\eqalign{F_j&=
-T\int {d\theta\over 2\pi}{1\over\cosh(\theta-\ln T_B/T)}
\ln\left(1+e^{-\epsilon_{j}}\right)\cr
F_{s-1}&=
-2T\int {d\theta\over 2\pi}{1\over\cosh(\theta-\ln T_B/T)}
\ln\left(1+e^{-\epsilon_{s-1}}\right).\cr}}
where $\epsilon_+=\epsilon_-\equiv\epsilon_{s-1}$.
Even though the BSG problem is integrable in this regime, applying
the TBA is  difficult technically since now
both the bulk and the boundary scattering matrices are not diagonal.
We will use analytic continuation again to provide the BSG free
energy.

Defining this time
$$Y_j(\alpha)=e^{-\epsilon_j(\alpha)},$$
we see right away that
for $j=2\dots s-2$,
\eqn\analogi{\eqalign{Y_1=&e^{-F_2/T}e^{-e^\alpha}\cr
Y_j=&e^{-F_{j+1}/T}e^{-F_{j-1}/T}\cr
Y_{s-1}=&e^{-F_{s-2}/T},\cr}}
analogous to \magicii.
Arguments identical to those in the attractive case
require that $Y(\alpha+is\pi)=Y(\alpha)$ \periodtba. Thus $Y$ can
be expressed an
analytic power series in
$x^2\propto e^{2\alpha/s} =(T_B/ T)^{2(1-g)}$
as before.
Therefore, \analogi\ indicates that
for $j$ even, $F_j+e^\alpha$ is a power series in $x^2$.
When $j$ is odd {\it and} $s$ is odd,
$F_j$ is a power series in $x^2$ as well, but
for $s$ even and $j$ odd, any other term is allowed as well.
In fact there  is a log term as  discussed in the last subsection.

We can find the log terms at $g=1-1/(2n)$
(i.e.\ $s$ even) directly from
the TBA, by plugging the power
series expansion for $Y$ into \sumfimp. For example, for $s=4$
($g=3/4$), $Y_1(\alpha)= 3 + a e^{\alpha/2} +b e^{\alpha}$.
Then, we see that
$$F_1/T + \ln 2 -f_2 x^2 =
-\int {d\theta\over 2\pi}{1\over\cosh(\theta-\ln T_B/T)}
\left[\ln\left(1+Y_1(\theta)\right) -\ln 4 - {a\over 4}
e^{\theta/2}\right].$$
For $T_B/T$ small, this is approximately
$$\eqalign{&-\int {d\theta\over 2\pi}{1\over\cosh(\theta-\ln T_B/T)}
\ln\left({4+a e^{\theta/2}+(b-a^2/8)e^\theta
\over 4 + a e^{\theta/2}}\right)\cr
\approx & -\int {d\theta\over 2\pi}{1\over\cosh(\theta-\ln T_B/T)}
{(b-a^2/8)e^\theta\over 4 + a e^{\theta/2}}\cr
= & {4T_B\over 2\pi T}(b-{a^2\over 8})
\int_0^\infty {du}{u^3\over 1 +u^4}
{1\over 4 + a \sqrt{T_B\over T}u}\cr
\approx & {b-a^2/8\over 4\pi} {T_B\over T} \ln {T_B\over T}.\cr}
$$
One can in fact verify using
functional relations analogous to those above and in \FLSjack\ that
$b-a^2/8=-2$,
so the coefficient is indeed $-4r_4=-T_B/(2\pi T)$ as shown in
the previous subsection.

We can now derive the analogs of the fusion relations
\fusion. We define
$$\eqalign{\tilde Z_j(x)&=e^{-F_j(\alpha)/T}\quad \qquad j\
\hbox{odd}\cr
\tilde Z_j(x)&=e^{-F_j(\alpha)/T}e^{-e^\alpha}\qquad j\
\hbox{even}.\cr}$$
Using \sident, we have
\eqn\analogii{\tilde Z_j((-q)^{1/2}x) \tilde Z_{j}((-q)^{-1/2}x)
 = 1+Y_j =1+ \tilde Z_{j-1}(x) \tilde Z_{j+1}(x)}
analogous to \zy\ and \trans\ in the attractive regime.
The crucial difference is that $q$ has been replaced by $-q^{-1}$.
The $\tilde Z_j(x)$ are analytic functions of $x^2$ for $j$ odd,
but for $j$ even they include the log term, so implicit
in this equation is the prescription $-i\pi < Im \ln y < i\pi$.
Because the analytic continuation of $Z_j$ should still satisfy
the fusion relation \trans, not all of the $\tilde Z_j$ can be
the analytic continuation
of the $Z_j$ from the attractive regime to the repulsive regime.
However, notice that if we make the identification
$$\eqalign{\ln \tilde Z_j(x)&=\ln Z_j(x) +{T_B\over T}
{\sin j\pi (s-1)/2 \over \cos\pi(s-1)/2}\qquad j\ \hbox{odd}\cr
 \ln \tilde Z_j(x)&=\ln Z_j(ix) +{T_B\over T}
{\sin j\pi (s-1)/2\over \cos\pi(s-1)/2}\qquad j\ \hbox{even}\cr}
$$
then the $Z_j$ satisfy the fusion relations \trans.
The shift as before cancels the pole in $\ln Z_j$ for $j$ odd.
Since $Z_j$ with $j$ even is a power series in $x^2$, the effect
of the argument $ix$ is to  flip the sign of every other term.
This is merely a matter of convention. With our choice $q=e^{i\pi
g}$, $q=1$ for the classical limit $g\to 0$
while $q=-1$ at the $SU(2)$ point
$g=1$. Representations of  $SU(2)_q$ and $SU(2)_{-q^{-1}}$ are
identical for $j$ odd but they differ by a  factor of $i$ in the
matrix
elements of $S_\pm$ for $j$ even.
The coupling renormalization for $j$ even would disappear if we chose
to change the quantum group conventions.

We can now find $Z_{BSG}$ by analytically continuing the functional
relation \fusiongoodie.
Since this relation involves only $q$ and the
functions are series in $x^2$, we can replace $q$ by $-q^{-1}$.
With this replacement, all functional relations derived in
in sect.\ 2.3\ apply to the repulsive regime with $t$ replaced
by $s$. In particular, we showed that
$2F_{BSG}(\xi x)=F_{t-1}(x)$ implies \fusiongoodie.
Since the relation \fusiongoodie\ for $g\ne 1$ determines all of the
BSG
coefficients $I_{2n}$ uniquely in terms of the spin-$1/2$ ones
$Q_{2n}$,
given a $Z_1$, it determines $Z_{BSG}$ uniquely.
Therefore, we can reverse the argument in sect.\ 2.3, and say that
given \fusiongoodie, we must have
\eqn\bsgkon{2F_{BSG}(\xi x)=F_{s-1}(x)}
where $\xi=i/(q-q^{-1})$ as before.
A similar result (missing the crucial factor of 2 and without
specifying
the $\xi$) was conjectured in \tsvel. We emphasize that this
relation is only true for $s$ integer. The relation \fusiongoodie\
of course is true for any value of $g$. However, the exact result
at integer $s$ is very useful, allowing us for example to find
the conductance in the BSG model exactly
at these values, without any analytic continuation.

We have checked the result \bsgkon\ numerically at
$s=3$ (again by comparing analytic
continuation results to TBA ones) and find good agreement.
The relation \bsgnorm\ still holds in the repulsive regime
(the derivation of \FLSbig\ holds for all $g$), and we confirm
also the value of $\xi$.

\newsec{Non-zero voltage}

We now extend the results of sect.\ 2 to allow for non-zero voltage
in
the BSG problem and non-zero
magnetic field in the Kondo model.
For the Kondo problem, the TBA analysis is easily extended
to non-zero magnetic field \refs{\AFL,\TW}. The analysis
is straightforward because the magnetic field couples to
a conserved charge, the $z$-component of the spin. Since
the charge commutes with the Hamiltonian, the same diagonalization
applies even with a magnetic field. However, in the BSG
problem the voltage violates the charge conservation. Indeed, this
is responsible for the charge tunneling in the Luttinger liquid
with an impurity.
In the presence of a voltage in
the BSG model, current flows,
so we can compute transport properties using a kinetic
equation \refs{\FLS,
\FLSjack,\FLSbig,\FLSnoise}.
However, this is not useful in finding the free energy at
non-zero voltage. Therefore one can use the series expansion
as above, or as below,
we will infer the free energy by utilizing a (well-checked)
conjecture. As a byproduct, we will also find some more information
about the transport properties. In particular, we conjecture
relations for the conductance good for all values of $g$,
generalizing the results of \refs{\FLS, \FLSjack,\FLSbig}.

The partition
functions can be expanded in powers of $x$ as before:
\eqn\expandv{\eqalign{Z_1(x,p)&=2\cos p\pi+\sum_{n=1}^\infty
x^{2n} Q_{2n}(p)\cr
Z_{BSG}(x,p)&=1+\sum_{n=1}^\infty
x^{2n} I_{2n}(p),}}
where
$$\eqalign{
Q_n(p)&=\int_{0}^{2\pi}du_1 \ldots\int_0^{u_n}du_n'\
{\cal I}_{2n}(\{u_i\},\{u_i'\})
2\cos p\left(\pi+\sum_i (u'_i-u_i)\right)\cr
I_{2n}(p)&={1\over (n!)^2}
\int_0^{2\pi}du_1\ldots  \int_0^{2\pi} du'_n\
{\cal I}_{2n}(\{u_i\},\{u_i'\})
\exp\left( ip\sum_i (u_i-u'_i)\right).\cr}$$
where $p=igV/2\pi T$.
In \FLSjack\ exact series expressions for
the $I_{2n}(p)$ were found for integer $p$:
\eqn\newIV{I_{2n}(p)= {1\over \Gamma(g)^{2n}}
\sum_{\bf m}
\prod_{i=1}^n {\Gamma\left[m_i+g(n-i+1)\right]
\Gamma\left[p+m_i+g(n-i+1)\right]
\over\Gamma\left[m_i+1+g(n-i)\right]
\Gamma\left[p+m_i+1+g(n-i)\right]} .
}
where ${\bf m}$ is defined as in \newIII.
As before, this series converges only for $g<1/2$.

Even though these results apply formally only to $p$ integer,
the above formula can be applied to all $p$ in the complex plane.
To prove that this is the unique analytic continuation, we
need to assume analyticity as $p\to\infty$. We know that
that $I_{2n}(p)/T^{2n(1-g)}$ is analytic as $p\to i \infty$
(where $T\to 0$) from the analysis of \FLSbig, and we assume  this
applies
at real $p$ as well. This assumption has been checked with TBA
results below. The series for $I_2$ can be summed as before:
\eqn\newitwo{\eqalign{I_2(p)=&\sum_{m_1=0}^\infty
{\Gamma(g+m_1) \Gamma(g+
\lambda_1+p)\over \Gamma^2(g)\Gamma(1+m_1)
\Gamma(1+m_1+p)}\cr
&={\sin{\pi g}\ \Gamma(1-2g)\over
\sin\pi (g+p)\Gamma(1-g+p)\Gamma(1-g-p)}.\cr}}
One can indeed check that this has the appropriate limit as
$p\to i\infty$ to reproduce the zero-temperature coefficient
of \FLSbig, giving support to our analyticity assumption.
Notice that the continuation of $I_{2n}(p)$ is not even in $p$,
and that $Z_{BSG}(p)$ is real only for $p$ integer.
The true partition function can be defined
by non-equilibrium methods like the Keldysh formalism.
However, we will see that observable quantities like the conductance
are given in terms of $Z_{BSG}(p)$.

The analysis of sect.\ 2.2 can be repeated for $p$ non-zero.
The fusion relations \fusion\ apply without modification for
spin-$j/2$ representations. The $k$-dimensional cyclic
representation when $q^k=\pm 1$ also obeys the same relation,
but to find the BSG free energy, there is a subtlety.
We first note that when $p\neq 0$ we have
\eqn\usefuli{Z_{\delta}(x,p)\approx
\left(\sum_{j=0}^{k-1} e^{i\pi jp}\right)
Z_{BSG}\left({Cq^{1/2}\over
q-q^{-1}}x,p\right),\qquad q^\delta=C>>1}
when fundamental
set of $S_z$ values is taken to be $0,\ldots, k-1$
(recall that periodic representations are invariant
under overall shifts of $S_z$). When one fuses such a
spin-$\delta$ representation with a
spin $1/2$ representation,
the last relation in \fusion\ still applies,
but the new cyclic representations have
shifted fundamental sets, $1,\ldots,k$  and $-1, \ldots,k-2$,
respectively. Therefore the relation between $Z_1$ and $Z_{BSG}$
is slightly modified:
\eqn\pfusiongoodie{Z_1[(q-q^{-1})x,p]={e^{i\pi p}Z_{BSG}(qx,p)
+e^{-i\pi p}Z_{BSG}(q^{-1}x,p)\over Z_{BSG}(x,p)}.}
Plugging in the power series expansions into \pfusiongoodie\ gives
the $I_{2n}(p)$ in terms of the $Q_{2n}(p)$. For
example,
$$I_2(p)={\sin{g \pi}\over\sin\pi\left(g
+p\right)}Q_2(p)$$
By direct, straightforward integration, one can check that
$$Q_2(p)={\Gamma(1-{2g})\over\Gamma\left(1-g+p\right)
\Gamma\left(1-g-p\right)},$$
in agreement with \newitwo.
At next order, we find
similarly
\eqn\resii{4\sin^3(\pi g) Q_4(p)=
-{2\cos\pi g\sin[\pi(2g-p)]} I_4(p)+
\sin[\pi(g-p)]I_2(p)^2.}
generalizing the $p=0$ relations in \FSkon. This
 relation \pfusiongoodie\ has been checked,
again by numerical determination of the TBA results
for the Kondo model at non-zero magnetic field.

The functional relation \pfusiongoodie\ implies even more than
the Kondo partition function. For example, we know on physical
grounds (and from the definition \expandv) that the Kondo partition
function is even in $p$. Combining this with
\pfusiongoodie\ then yields a non-trivial functional
relation for $Z_{BSG}$:
\eqn\willitend{\eqalign{e^{i\pi p} &\left[Z_{BSG}(qx,p)Z_{BSG}(x,-p)
-Z_{BSG}(q^{-1}x,-p)Z_{BSG}(x,p)\right]=\cr
&\qquad e^{-i\pi p} \left[Z_{BSG}(qx,-p)Z_{BSG}(x,p)
-Z_{BSG}(q^{-1}x,p)Z_{BSG}(x,-p)\right].\cr}}
Plugging in the perturbative expansion gives the values
of the coefficients $I_{2n}(-p)$ in term of the $I_{2n}(p)$:
$$\eqalign{&I_{2}(p)\sin[\pi (g+p)]=I_2(-p)\sin[\pi (g-p)]\cr
&I_{4}(p)\sin[\pi (2g+p)] - I_4(-p)\sin[\pi (2g-p)]= -\sin[\pi p]
I_2(p)I_2(-p),\cr}$$
for example.

In \FLSjack, we made and checked a conjecture which related the
linear-response conductance
directly to the partition function.
Our conjecture at general $\mu=V/2T$ and $g$ is
\eqn\cond{G(x,{V\over 2T})={g}-{ig\pi x\over 2}
{\partial\over \partial(V/2T)}{\partial\over \partial x}
\ln \left({Z_{BSG}(x,iV/2\pi tT)\over Z_{BSG}(x,-iV/2\pi
tT)}\right).}
We have checked this generalized expression numerically as well,
by comparing it to the conductance from the Boltzmann equations in
\refs{\FLS, \FLSbig} and below.

As at $V=0$, much is learned by comparing the perturbative
and TBA analyses. The TBA equations
\summeqatt\ at $g=1/t$ are modified slightly in the
presence of a finite voltage, yielding \refs{\AFL,\TW}
\eqn\summeqattbis{\epsilon_j=\sum_k N_{jk}\
 s_{t-1}*\ln\left(1+e^{\mu_k }e^{\epsilon_k}\right).}
where the incidence diagram is as in sect.\ 2. A magnetic
field corresponds to the chemical
potentials
$\mu_{\pm}=\pm \mu$, $\mu_k=0$ otherwise and $\mu\equiv V/2T$.
The two end nodes of the incidence diagram now play a
different role, but we still
 have $\epsilon_+=\epsilon_-\equiv\epsilon_{t-1}$. The relation
$$
\ln Z_j(x,\mu)=s_{t-1}*\ln\left(1+e^{\epsilon_j}\right),\quad
j=1,\ldots,t-2
$$
still holds. One also has
\eqn\eqi{\ln Z_{t-1}(x,\mu)=
s_{t-1}*\left[\ln\left(1+e^{\mu}e^{\epsilon_{t-1}}\right)
+\ln\left(1+e^{-\mu}e^{\epsilon_{t-1}}\right)\right].}
We then {\bf define}
\eqn\convdefs{\ln  Z_{\pm}(x,\mu)=
-{1\over 2}\ln\left[1+e^{\mu}{\sinh [(t-1)\mu /t]\over
\sinh [\mu/t]}\right]+s_{t-1}*
\ln\left(1+e^{\pm \mu}e^{\epsilon_{t-1}}\right).}
Using the same identity \sident\ as for $\mu=0$ yields
\eqn\rels{\eqalign{Z_\pm(q^{1/2}x,\mu)Z_\pm(q^{-1/2}x,\mu)&=
\left[1+e^{\pm \mu}{\sinh [(t-1)\mu/t]\over
\sinh [\mu/t]}\right]^{-1}(1+Y_\pm),\cr
Z_{t-1}&={\sinh \mu\over\sinh [\mu/t]}\quad Z_+Z_-.\cr}}
where
$$Y_\pm=e^{\pm \mu}e^{\epsilon_{t-1}}.$$
Similarly, relation \moremagic\ becomes
\eqn\relsii{Z_t(x,\mu)=Z_{t-2}(x,\mu)+2\cosh \mu,}
since the two additional states in the spin $t$ representation
 have third component of the spin equal to $\pm t$. The
fusion identites like \trans\ carry over to the case of finite
voltage.
Following the
 same arguments as for $\mu=0$ one finds then
\eqn\newnice{Z_1[(q-q^{-1})x,\mu]=e^{-\mu/t}{Z_+(qx,\mu)\over
 Z_-(x,\mu)}
+e^{\mu/t}{Z_-(q^{-1}x,\mu)\over Z_+(x,\mu)}.}
where we remind the reader that $\mu=V/2T$.

The functions $Z_\pm$ are not obviously related to any Kondo-type
integrals. Since there are technical obstacles to directly
calculating
the boundary sine-Gordon partition function in the presence of a
voltage,
we proceed using the algebraic approach.
Comparing \newnice\ and \pfusiongoodie\
suggests  the functional relations
\eqn\functnice{{Z_{BSG}(q^{1/2}x,i\mu/\pi t)\over
Z_{BSG}(q^{-1/2}x,i\mu/\pi t)}=
{Z_+(q^{1/2}x,\mu)\over Z_-(q^{-1/2}x,\mu)},}
together with
\eqn\evenmore{Z_{BSG}(x,{i\mu\over \pi t})
Z_{BSG}(x,-{i\mu\over \pi t})=
{\sinh [\mu/t]\over\sinh \mu} Z_{t-1}(x,\mu).}
Here we traded the $p$ variable of \expandv\ for the $\mu$ variable.
It is very likely that \evenmore\  has an algebraic
origin. This is because the tensor product of two cyclic
representations
decomposes on pairs of (generally) indecomposable representations,
which in turn
are related with  the spin $t-1$ representation of vanishing
$q$-dimension. However,
we have failed in finding a complete  algebraic proof of \evenmore,
but we have checked it thoroughly by using the series expressions
above for the $Z_{BSG}$ and the numerical TBA results for $Z_{t-1}$.

We can check that \functnice\ is consistent
with the conjectured relation \cond\
between the partition function
and the conductance. Using the TBA and a kinetic equation,
the conductance
at integer $t=1/g$ is \refs{\FLS,\FLSbig}
\eqn\conductttt{G(x,{V\over 2T})
={T(t-1)\over 2}{d\over dV}\int
{d\theta\over\cosh^2(t-1)(\theta-\alpha)}
\ln\left({1+e^{V/2T}e^{-\epsilon_{t-1}}\over
1+e^{-V/2T}e^{-\epsilon_{t-1}}}\right),}
Using the identity
\eqn\sidentii{\lim_{x\to 0} \left[{1\over \cosh^2(\theta+i\pi/2 -x)}-
{1\over \cosh^2(\theta-i\pi/2 +x)}\right]= -2i\pi \delta'(\theta),}
it follows that
\eqn\resssi{G(q^{1/2}x,{V\over 2T})-G(q^{-1/2}x,{V\over 2T})
=-{i\pi x\over 2 t}
{\partial\over\partial x}{\partial\over\partial (V/2T)}
\ln\left[{1+e^{V/2T}e^{\epsilon_{t-1}}\over
 1+e^{-V/2T}e^{\epsilon_{t-1}}}\right].}
This allows a powerful check on the conjectures \cond\ and
\functnice,
because it also follows from subsituting \functnice\ into \cond
and using the definition of $Z_\pm$ \convdefs.
It would be nice to reverse the order of the proof and show that
\resssi\ (known to be true from the TBA) implies \cond\ and
\functnice. This cannot be done by substituting
the perturbative expansion because
the relation \resssi\ does not determine all of them uniquely;
the order $x^{jt}$ term vanishes on the left-hand side for any
integer $j$. However, it is conceivable that by exploiting additional
analyticity information that it could be proven along the lines
of \Tracy.

In the linear-response limit $V\to 0$,
we can recover another functional relation from \FLSjack.
In this limit we can
ignore the $V$ dependence of $Y_{t-1}$ because it is
a function of $V^2$. Using \canyoubelieveit, we recover
\eqn\pet{G(q^{1/2}x,0)-G(q^{-1/2}x,0)={i\pi g^2 x} {\partial
\over \partial x}{1\over Z_{BSG}(q^{1/2}x,0)Z_{BSG}(q^{-1/2}x,0)}.}
This formula is nice because it
no longer has any reference to the TBA
quantities $\epsilon$, so we expect it to hold for all $g$.

We have a formula \evenmore\ which relates the product
of $Z_{BSG}(V)$ and $Z_{BSG}(-V)$ to TBA quantities, and a formula
\cond\
which relates their ratio to TBA quantities by using \conductttt.
Therefore, we can infer
a complete expression for $Z_{BSG}(x,\mu)$ alone in
terms of the TBA quantities:
\eqn\zvzvzv{\eqalign{
\ln Z_{BSG}(x,{iV\over2\pi tT})=
{t-1\over 2\pi}&
\int {d\theta}\left\{ {e^{-(t-1)(\theta-\alpha)}-i
\over 1+e^{-2(t-1)(\theta-\alpha)}}
\ln\left( {1+e^{\epsilon_{t-1}} e^{-V/2T}\over
 1+e^{\epsilon_{t-1}(-\infty)}e^{-V/2T}}\right)
\right.\cr
&+\left.
{e^{-(t-1)(\theta-\alpha)}+
i\over 1+e^{-2(t-1)(\theta-\alpha)}}
\ln\left( {1+e^{\epsilon_{t-1}} e^{V/2T}\over
 1+e^{\epsilon_{t-1}(-\infty)}e^{V/2T}}\right)\right\},\cr}}
where
$$
e^{\epsilon_{t-1}(-\infty)}={\sinh [(t-1)\mu/t]\over \sinh[\mu/t]}.
$$
and $\alpha=\ln(T_B/T)$ and $\mu=V/2T$ as always; $x$ is related
to $\alpha$ via \bsgnorm.
It should be possible to derive this directly from
the TBA, but there are some  technical obstacles.

Extending this analysis to the repulsive regime is more difficult.
The reason is that for integer $s$ in $g=1-1/s$,  the value of $S_z$
at the top and bottom states of the spin-$s$ representation
still is $\pm s$,
{\bf not} ${s\over s-1}$ as would be needed to carry over the
algebra of the
attractive regime straighforwardly. This issue is related to the
$q$ versus  $-q^{-1}$ problem
we had to address in sect.\ 3.2. We find after some manipulation that
\eqn\wildguess{{Z_{BSG}(q^{1/2}x,i\mu/\pi s)
\over Z_{BSG}(q^{-1/2}x,i\mu/\pi s)}=
{Z_+(q^{1/2}x,\mu)\over Z_-(q^{-1/2}x,\mu)},}
as above, but
with however the new correspondence $\mu=(s-1){V\over 2T}$.

We can also propose a formula for the conductance in the repulsive
 regime. By using the conjectures \evenmore\ and \cond\ and the
identity \sident,
we now deduce
$$\eqalign{&G\left(q^{1/2}x,{V\over 2T}\right)-G\left
(q^{-1/2}x,{V\over 2T}\right)=\cr
&\qquad\quad -{i\pi x\over 2} \left(1-{1\over s}\right)
{\partial\over\partial x}{\partial\over\partial (V/2T)}
\ln\left[{1+e^{(s-1)V/2T}e^{-\epsilon_{s-1}}\over
 1+e^{-(s-1)V/2T}e^{-\epsilon_{s-1}}}\right],\cr}$$
from which we obtain
\eqn\repcond{\eqalign{&G\left(x,{V\over 2T}\right)={s-1\over 2}\int
d\theta
 {1\over\cosh^2(\theta-\ln T_B/T)}\cr
&\qquad\times{d\over d(V/T)}\left\{\ln\left[
{1+e^{(s-1)V/2T}e^{-\epsilon_{s-1}
(\theta)}\over
 1+e^{-(s-1)V/2T}e^{-\epsilon_{s-1}(\theta)}}\right]
-\ln\left[ {1+e^{(s-1)V/2T}e^{-\epsilon_{s-1}(\infty)}\over
 1+e^{-(s-1)V/2T}e^{-\epsilon_{s-1}(\infty)}}\right]\right\}.\cr}}
By generalizing the result of eqn.\ \Tzero\
to non-zero $V$, one can check that these formulas give the correct
limit $G(0,V/2T)=(1-1/s)$. In the limit of vanishing voltage
(linear response), one finds
\eqn\repcondi{G(x,0)={(s-1)^2\over 2}
\int d\theta {1\over\cosh^2(\theta-\ln T_B/T)}
\left({1\over 1+e^{\epsilon_{s-1}(\theta)}}-{1\over
1+e^{\epsilon_{s-1}(\infty)}}\right)
.}
This expression has been compared to real-time
Monte Carlo simulations in
\lem; the agreement is good. At $T=0$, it agrees with the
expression derived in \FLSbig.
Using the identity \sidentii\
with \repcondi\ and using the TBA
expression for the free energy from sect.\ 3.2
yields the functional relation
\pet\ in the repulsive regime, lending support to the conjecture
that \pet\ holds for all $g$.

\newsec{Conclusion}

It should be possible to use the identification of the boundary
sine-Gordon model with a Kondo type problem more completely than we
have done. One way of doing so would be to write and solve the Bethe
ansatz equations for an integrable system made of spins $1/2$ and a
cyclic impurity. Since \genactbdry\ conserves the charge, unlike
\brdsg, this would allow one to handle directly the BSG model with a
voltage, avoiding the lengthy series of functional identities of
section 4.

Another interesting direction is to try to continue the perturbative
Anderson-Yuval coefficients past $g=1$ into the irrelevant regime.
This can be done by our zeta-function trick of section 3: use the
explicit series expression to find out how the series is diverging,
and subtract and add the appropriate zeta function. This is
straightforward but tedious, so we have not completed this program.
For example, the duality
$g\to 1/g$ \KF\ should be explicitly observable. So far this duality
has been established only at vanishing temperature \FLSbig.
Observe however that such a way of handling irrelevant operators
does not involve any cut-off, and will not describe the
non-universal physics
depending on the cutoff (e.g. the dissipative quantum mechanics
in \ref\FZ{M.P.A. Fisher and W. Zwerger, Phys. Rev. B32 (1985)
6190.}).
However, our result does provide
the exciting prospect of
a well-controlled irrelevant perturbation theory,
defined by an analog of dimensional regularization.

\bigskip
\centerline{\bf Acknowledgements}

We thank S. Chakravarty,
R. Egger, M.P.A. Fisher, A. Ludwig, K. Leung, C. Mak, J. Rudnick, A.
and Al.
Zamolodchikov for useful discussions.
Some of these results
were obtained independently by V. Bazhanov, S. Lukyanov
and A.B. Zamolodchikov (unpublished). This work was supported by the
Packard Foundation, the National Young Investigator program
(NSF-PHY-9357207), the DOE (De-FG03-84ER40168) and
a Canadian NSERC postdoctoral fellowship (F.L.).

\appendix{A}{Deriving the $S$ matrix for boundary sine-Gordon}

The boundary $S$ matrix of the boundary sine-Gordon model
was found in \GZ\ by analyzing the most general solution
of the boundary Yang-Baxter equation. Here we show that
this form also follows
from the identification of the BSG model
with a cyclic-spin anisotropic Kondo model.

For the ordinary spin-$j/2$ Kondo model,
the $S$ matrix for a particle scattering off the
impurity is easy to obtain. Up to
an overall proportionality factor which follows
from crossing and unitarity,
it is simply
the standard Yang-Baxter solution for a spin $(j-1)/2$ and
a spin $1/2$, and a renormalized quantum group parameter
$q=e^{i\pi g/(1-g)}$ \pkondo.
By analogy, we expect the $S$ matrix for the cyclic spin case to
be given (up to the overall factor), by the Yang-Baxter solution
for a cyclic spin and a spin $1/2$. This $R$ matrix is the object
studied long ago
\ref\skly{E.K. Sklyanin, J. Phys. A21 (1988) 2375.}.
It is conveniently written
as a matrix in $\Pi_1\otimes\Pi_\delta$
\eqn\skl{\tau=\left(\eqalign{w_0S_0+2w_3S_3&\quad  2w_1 S_-\cr
2w_1S_+ &\quad w_0S_0-2w_3S_3\cr}\right),}
where
\eqn\defspara{a=\sin(\gamma+u),b=\sin u,c=\sin\gamma,}
and
$$w_0={a+b\over 2},w_3={a-b\over 2},w_1={c\over 2},$$
and in the cyclic representation, $S_\pm$ act as indicated above,
while
$$\eqalign{
S_0|m>&={q^m+q^{-m}\over q^{1/2}+q^{-1/2}}\cr
S_3|m>&={q^m-q^{-m}\over 2(q^{1/2}-q^{-1/2})}.\cr}$$
Setting $u={\pi\over t}({1\over 2}-\delta)+v$ and taking
as above the limit $q^\delta=C>>1$, one finds the $R$ matrix
\eqn\limit{\eqalign{R_{+,m}^{+,m}&=e^{-iv}q^{-m}\cr
R_{+,m}^{-,m+1}&=q^{-m}\cr
R_{-,m}^{-,m}&=e^{-iv}q^m\cr
R_{-,m}^{+,m-1}&=q^m.\cr}}
The result of \GZ\ is that
the boundary sine-Gordon $S$ matrix provides a
solution of the {\bf boundary} Yang-Baxter equations
of the form
\eqn\bdrr{R=\left(\eqalign{e^{-iv}& \quad 1\cr
1&\quad  e^{-iv}\cr}\right),}
while \limit\ is a solution of the {\bf ordinary}
Yang-Baxter  equation. To map
the two, it is tempting to simply
forget the cyclic degrees of freedom, which
appear only as rapidity-independent phases. It is however not
totally possible, and this has to do with the difference
between BYB and YB even for massless particles. Indeed
in a massless theory the left-right scattering is
rapidity independent, but it might still involve some
phases. As such, \bdrr\ solves BYB but does {\bf not} solve YB,
because of the left-right scattering phases.
When considering YB, there are no left-right scattering phases,
so the equivalent terms are furnished  by the
cyclic degrees of freedom in \limit. This is illustrated
in figures 4 and 5.
The complete translation shows that,
if \limit\ satisfies YB, then \bdrr\ satisfies BYB indeed.
In the  figures $4a,b,c$ we consider a particular case of the YB
equation
involving the scattering of two spins $1/2$ and a ``cyclic'' spin.
The weight
of the first figure is $W_1=a(u-v)e^{-iu}e^{-2imt}$, the one of the
second figure $W_2=b(u-v)e^{-iu}e^{-i\pi m/t}e^{-i\pi (m+1)/t}$, and
the third $W_3=
ce^{-iv}e^{-2i\pi m/t}$. The fact that YB  holds means that
$W_1=W_2+W_3$.

In the figures $5a,b,c$
we consider a particular case of the BYB
equation
involving a spin $1/2$ bouncing off a boundary. The weight of the
first figure is $W'_1=e^{-iu}a_{LL}(u-v)a_{LR}(u-v)$, the weight of
the second figure
$W_2'=e^{-iu}b_{LL}(u-v)b_{LR}(u-v)$, and the weight of the third
$W_3'=e^{-iv}c_{RR}(u-v)a_{LR}(u-v)$. The fact that BYB holds means
$W_1'=W_2'+W_3'$, which one checks easily since the RR elements are
identical
to the ones in \defspara\ and
$a_{LR}=1,b_{LR}=e^{-i\gamma},c_{LR}=0$.
\vfill\eject
\listrefs
\fig{Configurations involved in checking YB equation with two
spin $1/2$ and a cyclic representation (see the appendix)
}{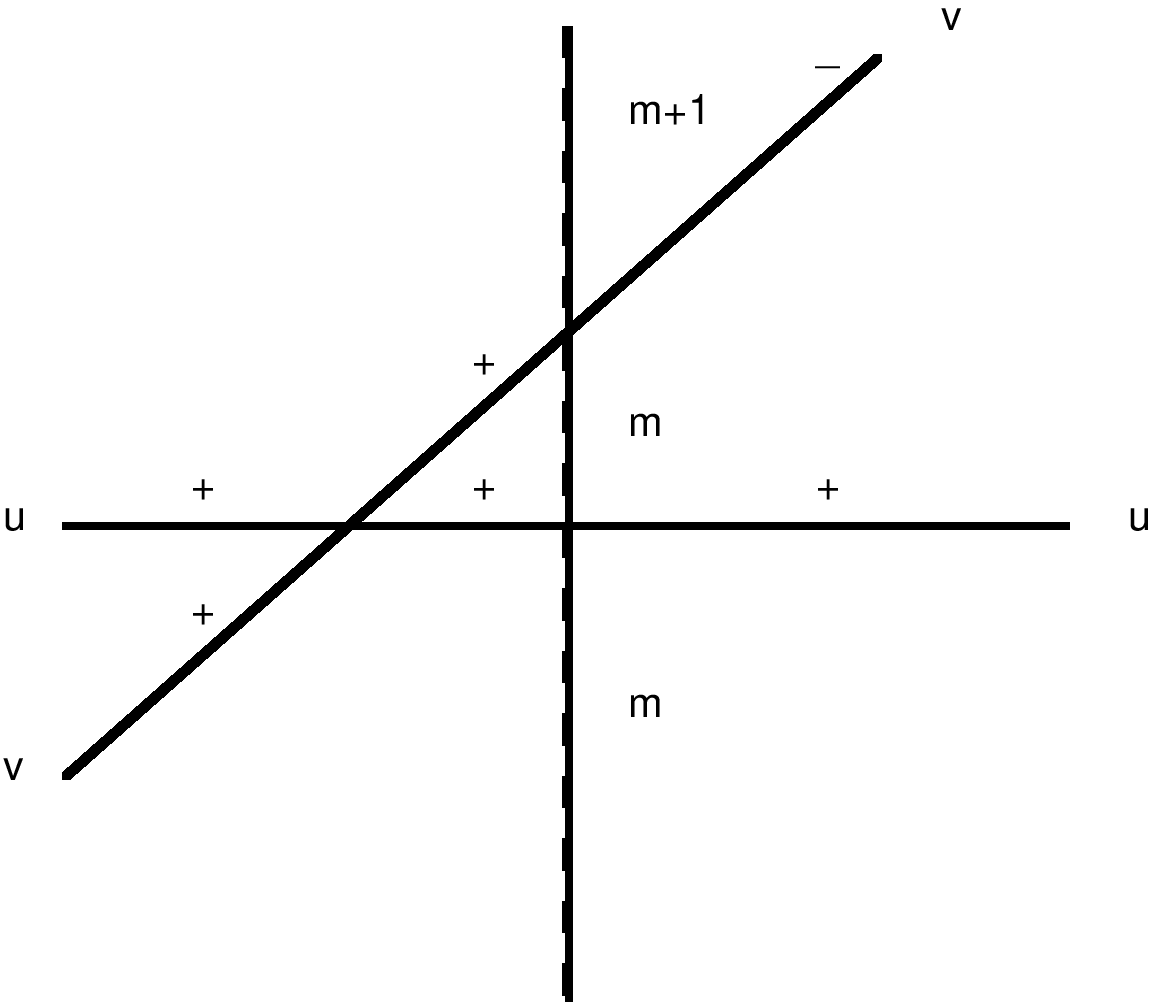}{10cm}{4a}
\figlabel\tabb
\fig{{\it idem}
}{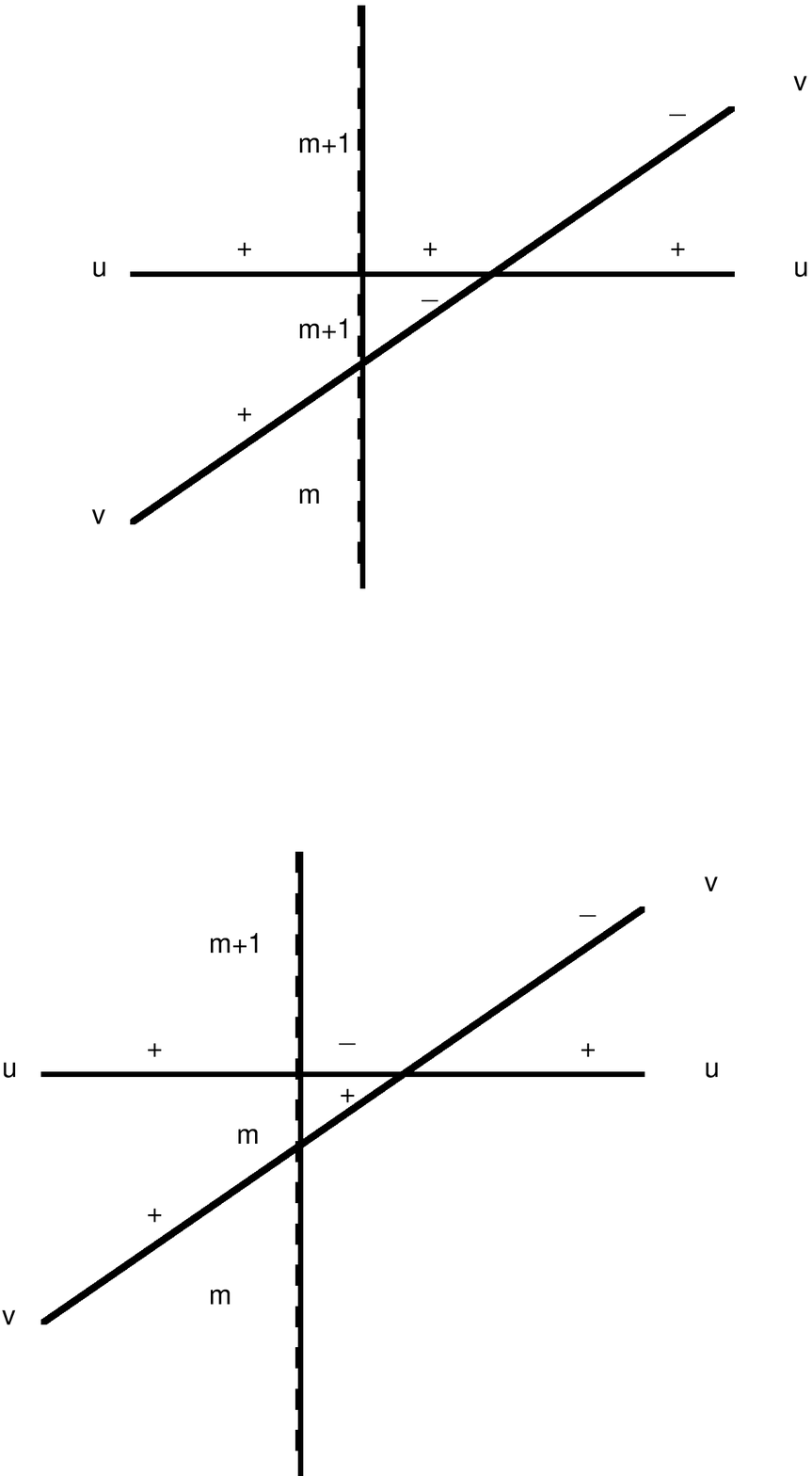}{10cm}{4b,c}
\figlabel\tabb
\fig{Configurations involved in checking BYB with no degree of
freedom at the boundary (see the apppendix)}{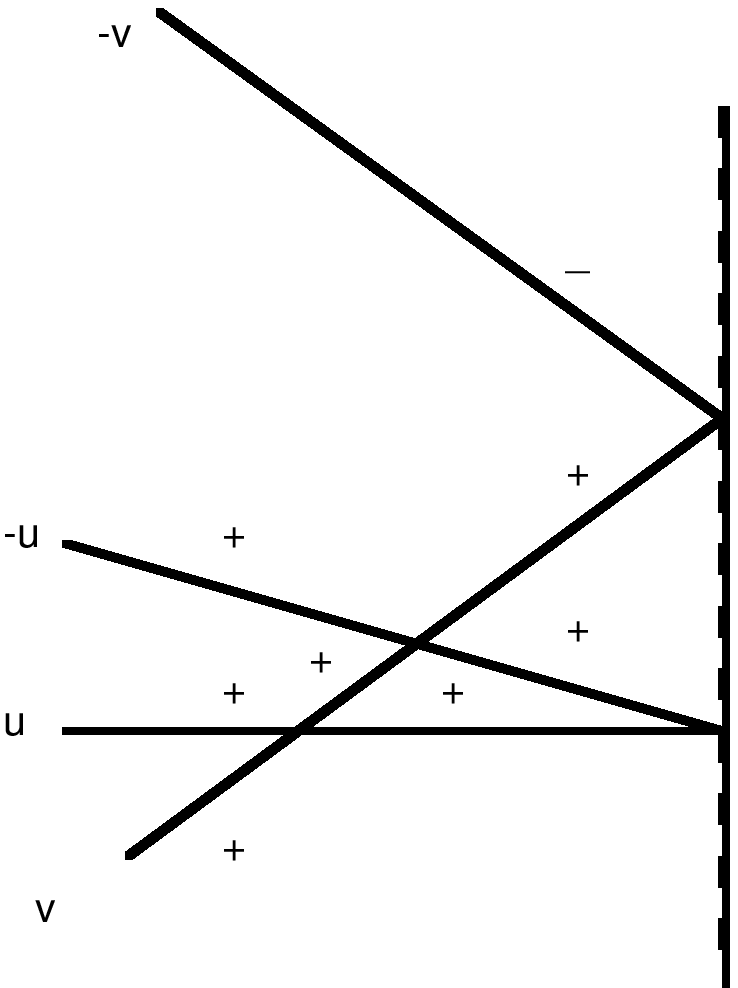}{8cm}{5a}
\figlabel\tabb
\fig{\it idem
}{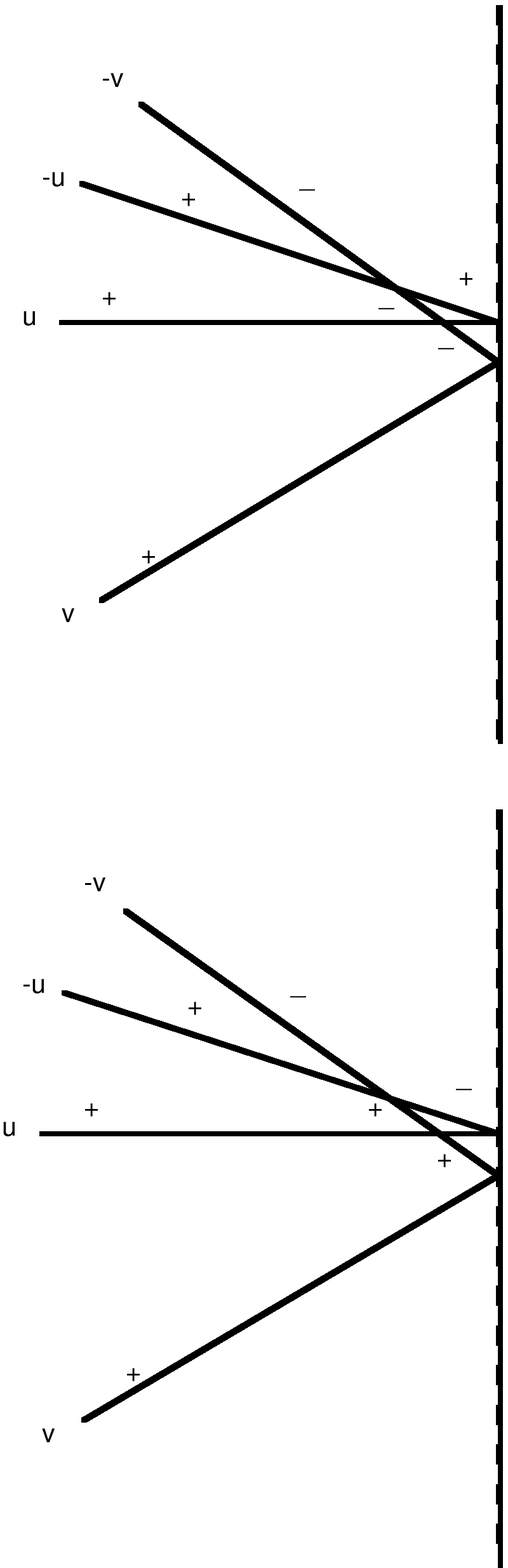}{6cm}{5b,c}
\figlabel\tabb
\bye